\documentclass[11pt]{article}
\usepackage{amsmath,amssymb}
\usepackage{latexsym}
\usepackage{graphicx}
\usepackage{amscd,color}
\usepackage{theorem}
\input xy
\xyoption{all}

\newcommand{\R}{{\mathbb{R}}}
\newcommand{\Z}{{\mathbb{Z}}}
\newcommand{\C}{{\mathbb{C}}}

\newcommand{\bea}{\begin{eqnarray}}
\newcommand{\eea}{\end{eqnarray}}

\newcommand{\nn}{\nonumber}

\newcommand{\bp}{\begin{pmatrix}}
\newcommand{\ep}{\end{pmatrix}}

\newcommand{\bps}{\begin{smallmatrix}}
\newcommand{\eps}{\end{smallmatrix}}

\newcommand{\ti}{\tilde}
\newcommand{\la}{\langle}
\newcommand{\ra}{\rangle}

\def \cA{{\cal A}}

\def \cH{{\cal H}}
\def \cM{{\cal M}}
\def \cMC{{\cal MC}}
\def \cO{{\cal O}}

\def \V{{\cal V}}

\def \f{{\frak f}}
\def \g{{\frak g}}
\def \l{{\frak l}}
\def \m{{\frak m}}
\def \n{{\frak n}}

\def \S{{\frak S}}

\def \raw{\rightarrow}

\def \oraw{\overrightarrow}

\def \deg{\mathrm{deg}}

\def \Hom{\mathrm{Hom}}
\def \Coder{\mathrm{Coder}}

\def\ott{\otimes}

\def \cb{{\bar c}}
\def \ob{{\bar o}}

\def \half{\frac{1}{2}}
\def \ov#1{\frac{1}{#1}}

\def \mb{\bar{m}}

\def \bpart{\bar{\partial}}

\def \flpartial#1{\frac{\overleftarrow{\partial}}{\partial #1}}
\def \frpartial#1{\frac{\overrightarrow{\partial}}{\partial #1}}

\def \({\left(}
\def \){\right)}

\def \0{{\bf 0}}
\def \1{{\bf 1}}

\def \eb{{\bf e}}

\def \-{-\hspace*{-0.2cm}-}
\def \ie{{\it i.e.}\ }

\def \Nsddata#1#2#3#4#5{
(\xymatrix{{#1}\  \ar@<0.5ex>[r]^{{#2}} & \ {#4}
\ar@<0.5ex>[l]^{{#3}}} ,#5) }

{
 \newtheorem{thm}{Theorem}[section]
 
 \newtheorem{lem}[thm]{Lemma}
 \newtheorem{cor}[thm]{Corollary}

\theorembodyfont{\rmfamily}
 \newtheorem{defn}[thm]{Definition}%[section]
 \newtheorem{rem}[thm]{Remark}%[section]
}

\numberwithin{equation}{section}

\begin{document}

\begin{titlepage}
\thispagestyle{empty}
\begin{flushleft}
\hfill YITP-05-62\\
\hfill October, 2005\\ 

\end{flushleft}

%\vskip 1.5 cm
\vskip 0.5cm

\begin{center}
\noindent{\Large \textbf{Open-closed homotopy algebra 
in mathematical physics}}\\

\renewcommand{\thefootnote}{\fnsymbol{footnote}}

%\vskip 2cm
\vskip 0.7cm

{\large 
\noindent{ \bigskip }\\

\it
Hiroshige Kajiura${}^{*}$ and Jim Stasheff${}^{\dag}$\\
%\noindent{\smallskip  }\\
\vskip 0.5cm

%\noindent{\smallskip  }\\

${}^{*}$
Yukawa Institute for Theoretical Physics, Kyoto University \\
Kyoto 606-8502, Japan \\
e-mail: kajiura@yukawa.kyoto-u.ac.jp\\

\noindent{\smallskip  }

${}^{\dag}$
Department of Mathematics, University of Pennsylvania \\
Philadelphia, PA 19104-6395, USA \\
e-mail: jds@math.upenn.edu\\

}

\end{center}
\begin{abstract}
In this paper we discuss various aspects of open-closed homotopy 
algebras (OCHAs) presented in our previous paper, 
inspired by Zwiebach's open-closed string field theory, 
but that first paper concentrated
on the mathematical aspects. 
Here we show how an OCHA is obtained by extracting the tree part of 
Zwiebach's quantum open-closed string field theory. 
We clarify the explicit relation of an OCHA with 
Kontsevich's deformation quantization and with the B-models of 
homological mirror symmetry.
An explicit form of the minimal model for an OCHA is  
given as well as  its relation to the perturbative expansion 
of open-closed string field theory. 
We show that our open-closed homotopy algebra gives us
a general scheme for deformation of
open string structures ($A_\infty$-algebras)
by closed strings ($L_\infty$-algebras).
\end{abstract}

\renewcommand{\thefootnote}{}
\footnote{H.~K is supported by JSPS Research Fellowships 
for Young Scientists. 
J.~S. is supported in part by 
NSF grant FRG DMS-0139799 
and US-Czech Republic grant INT-0203119. 
}

\setcounter{footnote}{0}

\vfill

\end{titlepage}
\vfill
\setcounter{footnote}{0}
\renewcommand{\thefootnote}{\arabic{footnote}}
\newpage

\tableofcontents

\markboth{Open-closed homotopy algebra 
in mathematical physics, \qquad Kajiura-Stasheff}
{\scriptsize \em OPEN--CLOSED HOMOTOPY ALGEBRA IN 
MATHEMATICAL PHYSICS, \qquad KAJIURA--STASHEFF}
%{Hiroshige Kajiura and Jim Stasheff {\rm [\today]}}

 \section{Introduction}
\label{sec:1}

In this paper we discuss various aspects of open-closed homotopy 
algebras (OCHAs) 
defined in our previous paper \cite{OCHA}. 
They are a kind of homotopy algebra 
inspired by Zwiebach's classical open-closed string field theory \cite{Z2} 
and also related to the deformation quantization set-up by 
M.~Kontsevich \cite{Ko1}. 
In \cite{OCHA} 
we showed that an OCHA actually defines a homotopy invariant 
algebraic structure 
and also it gives us a general scheme for deformation of 
open string structures ($A_\infty$-algebras) 
by closed strings ($L_\infty$-algebras). 

As tree closed strings and open strings are related to 
the conformal plane $\C$ with punctures 
and the upper half plane $H$ with punctures on the boundary,
respectively, 
tree open-closed strings are related to 
the upper half plane $H$ with punctures both in the bulk and on the boundary, 
which appears recently 
in the context of deformation quantization \cite{Ko1}. 
In operad theory (see \cite{MSS}), 
the relevance of the little disk operad to closed string 
theory is known. 
The little interval operad and associahedra are  relevant to 
open string theory. 
The Swiss-cheese operad \cite{Vo}, that combines the little disk 
operad with the little interval operad, 
also is inspired by Kontsevich's approach to deformation quantization.
Our OCHA should be homotopy equivalent to a part of 
an algebra over the Swiss-cheese operad. 
It should be very interesting to investigate the remaining 
structures (see \cite{erich}, which is related to this direction). 
 
We first present the definition of OCHAs 
together with recalling two typical homotopy algebras, 
$A_\infty$-algebras and $L_\infty$-algebras, in section \ref{sec:oc}. 

In section \ref{sec:dual}, we give an alternate interpretation 
in terms of odd formal vector fields 
(often called homological vector fields) on a supermanifold, 
which we believe is a more acceptable description for physicists. 
 
The connection to classical open-closed string field theory by 
B.~Zwiebach \cite{Z2} is given in section \ref{sec:ocsft}.  
It is known that 
classical closed string field theory has an 
$L_\infty$-structure \cite{Z1,Sta1993,KSV}, and 
classical open string field theory 
has an $A_\infty$-structure  \cite{GZ,Z2,N,Ka}. 
We show that an OCHA is obtained by extracting the tree part of 
Zwiebach's quantum open-closed string field theory.  
Since in general homotopy algebras are something whose structures are 
governed by the underlying tree graph (operad) structure, 
the structures of quantum string field theories are 
something beyond the ordinary homotopy algebra 
(see loop homotopy algebras \cite{markl:loop} for quantum closed 
string field theories). 
Thus, we can say that OCHAs are the `maximal' homotopy algebraic 
structures which string field theories should have. Namely, 
\begin{equation*}
 \mbox{Quantum open-closed SFT}\supset 
 \mbox{OCHA} \supset L_\infty \oplus A_\infty\ .  
\end{equation*}

One of the key theorems in homotopy algebra is the minimal model
theorem which was proved for $A_\infty$-algebras by 
Kadeishvili \cite{kadei1}. 
It holds true also for $L_\infty$-algebras in a similar way, 
and in our previous paper \cite{OCHA} 
we stated the minimal model theorem holds for OCHAs, too. 
In section \ref{sec:MMth} we present an explicit way of 
constructing a minimal model for an OCHA, and 
explain its relation to the perturbative expansion of 
classical open-closed string field theory.

Section \ref{sec:def} is devoted to explaining some 
deformation theory aspect of OCHAs. 
An open-closed homotopy algebra consists of a direct sum of 
graded vector spaces $\cH =\cH_c\oplus\cH_o$. It has 
an $L_\infty$-structure on $\cH_c$ and reduces to 
an $A_\infty$-algebra if we set $\cH_c=0$. 
{}From such a viewpoint, an open-closed homotopy algebra 
gives a general scheme of 
deformation of the $A_\infty$-algebra by $\cH_c$, 
where the deformation space is parameterized by a moduli space 
of the $L_\infty$-algebra on $\cH_c$ \cite{OCHA}. 
In subsection \ref{ssec:def} 
we recall this fact in a more explicit way than \cite{OCHA}. 
After that, we explain the relation of this viewpoint to 
various aspects of string theory; 
Kontsevich's deformation quantization \cite{Ko1} 
in subsection \ref{ssec:K}, 
and open-closed B-models (cf. \cite{Hof2}) in subsection \ref{ssec:Bmodel}.

 \section{Open-closed homotopy algebra (OCHA)}
\label{sec:oc}

An open-closed homotopy algebra, as we proposed in our previous paper 
\cite{OCHA}, is a 
homotopy algebra which combines two typical homotopy algebras, 
an $A_\infty$-algebra and an $L_\infty$-algebra. 
There are various equivalent ways of defining/describing strong 
homotopy algebras. 
In this paper, we shall present them 
in terms of multi-variable operations in this section, and 
in section \ref{sec:dual} we shall reinterpret them 
in terms of the supermanifold description. 
For the equivalent coalgebra description and tree graph description, 
see \cite{OCHA}. Here we recall just enough so that this paper
can be read without having to read \cite{OCHA}.  
The reader familiar
with $A_\infty$-algebras and $L_\infty$-algebras can go directly 
to Definition \ref{defn:oc}.

We first begin with recalling $A_\infty$-algebras and
$L_\infty$-algebras 
in subsection \ref{ssec:AandL}. 
The definition of open-closed homotopy algebras 
are given in subsection \ref{ssec:ocha}. 
In subsection \ref{ssec:cyclic} 
we define cyclic structures in open-closed homotopy algebras 
together with explaining some background of such structures. 

We restrict our arguments to the case that the characteristic of 
the field $k$ is zero. We further let $k=\C$ for simplicity. 

 \subsection{$A_\infty$-algebras and $L_\infty$-algebras}
\label{ssec:AandL}

\begin{defn}[$A_\infty$-algebra\ 
(strong homotopy associative algebra)\cite{Sta}]
Let $\cH_o$ be a $\Z$-graded vector space 
$\cH_o=\oplus_{r\in\Z}\cH_o^r$ and suppose that 
there exists a collection of degree one multi-linear maps 
\begin{equation*}
 \m:=\{m_k : (\cH_o)^{\otimes k}\raw\cH_o\}_{k\ge 1} \ .
\end{equation*}
$(\cH_o,\m)$ is called an {\em  $A_\infty$-algebra} when the multi-linear
maps $m_k$ satisfy the following relations 
\begin{equation}
\sum_{k+l=n+1}\sum_{j=0}^{k-1}
{(-1)^{o_1+\cdots+o_j}
 m_k(o_1,\cdots,o_j,m_l(o_{j+1},\cdots,o_{j+l}),
 o_{j+l+1},\cdots,o_n)}=0\ 
 \label{Ainfty}
\end{equation}
for $n\ge 1$, 
where $o_i$ on $(-1)$ denotes the degree of $o_i$. 
A {\em weak $A_\infty$-algebra} $(\cH_o,\m)$ consists of a 
collection of degree one multi-linear maps 
$\m:=\{m_k : (\cH_o)^{\otimes k}\raw\cH_o\}_{k\ge 0}$ 
satisfying the corresponding relations: 
\begin{equation*}
\sum_{k+l=n+1}\sum_{j=0}^{k-1}
{(-1)^{o_1+\cdots+o_j}
 m_k(o_1,\cdots,o_j,m_l(o_{j+1},\cdots,o_{j+l}),
 o_{j+l+1},\cdots,o_n)}=0\ 
\end{equation*}
for $n\ge 0$.  
 \label{defn:Ainfty}
\end{defn}
\begin{rem}
The definition above is different from the original one \cite{Sta} 
in the definition of the degree of the multi-linear maps $m_k$. 
Both are in fact equivalent and related by  
{\it suspension} \cite{GJ,MSS}.
In \cite{Sta}, the $m_k$ are multi-linear maps on $\downarrow \cH_o$ where
$(\downarrow \cH_o)^{r+1} = \cH_o^r$; in algebraic topology 
the desuspension is denoted by $\downarrow$, which is equivalent to 
$[-1]$ in the algebraic geometry tradition:$\downarrow \cH_o = \cH_o[-1]$. 
Since it might be more familiar also in mathematical physics 
as in section \ref{sec:def}, in this paper 
we denote the suspension and desuspension by $[1]$ and $[-1]$, 
respectively. 
 \label{rem:sus}
\end{rem}
For an $A_\infty$-algebra $(\cH_o,\m)$ (in the case $m_0=0$), 
the first three relations of the $A_\infty$-condition (\ref{Ainfty}) are :
\begin{align*}
0 &=m_1^2\ ,\\
0 &=m_1(m_2(o_1,o_2))+m_2(m_1(o_1),o_2)
+(-1)^{o_1}m_2(o_1,m_1(o_2))\ ,\\
0 &=m_1(m_3(o_1,o_2,o_3))+
m_3(m_1(o_1),o_2,o_3)+(-1)^{o_1}m_3(o_1,m_1(o_2),o_3)
+(-1)^{o_1+o_2}m_3(o_1,o_2,m_1(o_3))\\
&\qquad 
+m_2(m_2(o_1,o_2),o_3)+(-1)^{o_1}m_2(o_1,m_2(o_2,o_3))\ .
\end{align*}
The first equation, in the physics terminology, says $m_1$ is nilpotent; 
$(\cH_o, m_1)$ defines a complex on the $\Z$-graded vector space $\cH_o$. 
The second equation says the differential $m_1$ satisfies the
Leibniz rule for the product $m_2$. 
The third equation means the product $m_2$ is associative up to the 
term including $m_3$. 
Thus, a differential graded algebra (DGA) is described as 
an $A_\infty$-algebra on $\downarrow\cH_o=\cH_o[-1]$ with 
a differential $m_1$, a product $m_2$, and $m_3=m_4=\cdots=0$. 
\begin{defn}[$A_\infty$-morphism]
For two $A_\infty$-algebras $(\cH_o, \m)$ and $(\cH_o',\m')$,
suppose that there exists a collection of
degree zero (degree preserving) multi-linear maps
\begin{equation*}
 f_k: \cH_o^{\otimes k}\raw \cH_o'\ ,\qquad k\geq 1\ .
\end{equation*}
The collection
$\{f_k\}_{k\ge 1}:(\cH_o,\m)\raw (\cH_o',\m')$ is called an {\em
$A_\infty$-morphism} iff it satisfies the following relations:
\begin{equation}
 \begin{split}
& \sum_{1\leq k_1<k_2\dots <k_i=n}\
{m'_i(f_{k_1}(o_1,\cdots,o_{k_1}),
f_{k_2-k_1}(o_{k_1+1},\cdots,o_{k_2})\cdots f_{n-k_{i-1}}
(o_{k_{i-1}+1},\cdots,o_n))}\\
&\qquad=\sum_{k+l=n+1}\sum_{j=0}^{k-1}{(-1)^{o_1+\dots +o_j}
f_k(o_1,\cdots,o_j,m_l(o_{j+1},\cdots,o_{j+l}),
o_{j+l+1},\cdots,o_n)}\
\end{split}
\label{amorphism}
\end{equation}
for $n\ge 1$.
If $(\cH_o, \m)$ and $(\cH_o',\m')$ are {\em weak} $A_\infty$-algebras, 
then a {\em weak} $A_\infty$-morphism consists of multi-linear maps
$\{f_k\}_{k\ge 0}$, where $f_0:\C\raw A'$, 
satisfying the above conditions and in addition:
$$
 f_1\circ m_0 = \sum m'_k(f_0,\cdots,f_0)\ .
$$
 \label{defn:Ainftymorp}
\end{defn}
As an $A_\infty$-algebra can be thought of as a generalization of a
differential graded algebra 
(DGA), an $L_\infty$-algebra is a generalization of a differential graded 
Lie algebra (DGLA). 
As  ordinary associative and Lie algebras are related by skew-symmetrization
and the universal enveloping construction, there are corresponding relations
for $A_\infty$-algebras and $L_\infty$-algebras
\cite{lada-markl}. 
\begin{defn}[Graded symmetry]
A {\em graded symmetric multi-linear map}  of a
graded vector space $V$ to itself is a linear map
$f:V^{\otimes n}\to V$ such that, for any $c_i\in V$ $1\le i\le n$ and
any $\sigma\in\S_n$ (the permutation group of $n$ elements), the relation
\begin{equation}
 f(c_1,\cdot\cdots,c_n)
 =(-1)^{\epsilon(\sigma)}f(c_{\sigma(1)},\cdots\cdot,c_{\sigma(n)})
 \label{gcomm}
\end{equation}
holds, where 
the sign  $(-1)^{\epsilon(\sigma)}$ is 
the Koszul sign of the permutation $\sigma$.
 \label{defn:gcomm}
\end{defn}
Also we adopt the convention that tensor products of functions or
operators
have the signs built in; e.g. $(f\ott g)(x\ott y) = (-1)^{g\cdot
x}f(x)\ott g(y).$ 
\begin{defn}[ $L_\infty$-algebra\ (strong homotopy Lie algebra)
\cite{LS}]
Let $\cH_c$ be a graded vector space and suppose that
a collection of degree one graded symmetric multi-linear maps
$\l:=\{l_k:\cH_c^{\otimes k}\raw\cH_c\}_{l\ge 0}$ is given. 
$(\cH_c,\l)$ is called a {\em weak  $L_\infty$-algebra} iff
the multi-linear maps satisfy the following relations :
\begin{equation}
\sum_{k+l=n+1}\sum_{\sigma\in\S_n}
\frac{(-1)^{\epsilon(\sigma)}}{l!(n-l)!}
l_k(l_l(c_{\sigma(1)},\cdots,c_{\sigma(l)}),
 c_{\sigma(l+1)},\cdots,c_{\sigma(n)})=0\
 \label{Linfty}
\end{equation}
for $n\ge 0$. 
If the relation is satisfied for $n\ge 1$ without the
additional map $l_0:\C\raw\cH_c^1\subset\cH_c$, 
then $(\cH_c,\l)$ is called an {\em $L_\infty$-algebra}.
 \label{defn:Linfty}
\end{defn}
\begin{rem}
$L_\infty$-algebras are usually defined in a similar but different
fashion,
where the summation for the permutation $\S_n$ in eq.(\ref{Linfty})
is replaced by the summation over the `unshuffle' permutations
(\ref{gcomm}).
This `unshuffled' description would enable us to drop all the
symmetrization factors in this paper. However, we take the one with all
the permutations
since it fits the dual description in the next section.
 \label{rem:unshuffle}
\end{rem}
For an $L_\infty$-algebra $(\cH_c,\l)$, 
the first three relations of the $L_\infty$-condition (\ref{Linfty}) are :
\begin{align*}
0 &=(l_1)^2\ ,\\
0 &=l_1(l_2(c_1,c_2))+l_2(l_1(c_1),c_2)
+(-1)^{c_1}l_2(c_1,l_1(c_2))\ ,\\
0 &=l_1(l_3(c_1,c_2,c_3))+
l_3(l_1(c_1),c_2,c_3)+(-1)^{c_1}l_3(c_1,l_1(c_2),c_3)
+(-1)^{c_1+c_2}l_3(c_1,c_2,l_1(c_3))\\
&\qquad 
+l_2(l_2(c_1,c_2),c_3)+(-1)^{c_1(c_2+c_3)}l_2(l_2(c_2,c_3),c_1)
+(-1)^{c_3(c_1+c_2)}l_2(l_2(c_3,c_1),c_2)\ .
\end{align*}
As in the case of an $A_\infty$-algebra, 
the first equation indicates that 
$(\cH_c, l_1)$ defines a complex, while, after a shift in grading,
the second equation implies the differential $l_1$ satisfies a Leibniz rule 
with respect to the Lie bracket $l_2$, and 
the third equation means the bracket $l_2$ satisfies the 
Jacobi-identity up to the terms including $l_3$. 
Thus, a differential graded Lie algebra (DGLA) is described as 
an $L_\infty$-algebra on $\downarrow\cH_c=\cH_c[-1]$ with 
a differential $l_1$, a Lie bracket $l_2$, and $l_3=l_4=\cdots=0$. 
\begin{defn}[$L_\infty$-morphism]
For two weak $L_\infty$-algebras $(\cH_c,\l)$ and $(\cH_c',\l')$,
suppose that there exists a collection of
degree zero (degree preserving) graded symmetric multi-linear maps
\begin{equation*}
 f_k: \cH_c^{\otimes k}\raw \cH_c'\ ,\qquad l\ge 0\ .
\end{equation*}
Here $f_0$ is a map from $\C$ to a degree zero sub-vector space of
$\cH_c$. The collection
$\{f_k\}_{k\ge 0}:(\cH_c,\l)\raw (\cH_c',\l')$ is called a {\em weak
$L_\infty$-morphism} iff it satisfies the following relations
\begin{equation}
 \begin{split}
&\sum_{k+l=n+1}\sum_{\sigma\in\S_n}
\frac{(-1)^{\epsilon(\sigma)}}{l!(n-l)!}
f_k(l_l\otimes \1_c^{\otimes n-l})
(c_{\sigma(1)},\cdots,c_{\sigma(n)})\\
&=\sum_{k_1+\cdots +k_j =n}\,
\sum_{\sigma\in\S_n}
\frac{(-1)^{\epsilon(\sigma)}}
{k_1!k_2!\cdots k_j!\cdot j!} l'_j(f_{k_1}\otimes
f_{k_2}\otimes \cdots \otimes f_{k_j})
(c_{\sigma(1)},\cdots,c_{\sigma(n)})
\end{split}
\label{Linftymorp}
\end{equation}
for $n\ge 0$. 
In particular, 
when $(\cH_c,\l)$ and $(\cH_c',\l')$ are $L_\infty$-algebras, 
a weak $L_\infty$-morphism $\{f_k\}_{k\ge 0}:(\cH_c,\l)\raw (\cH_c',\l')$ 
is called an {\em $L_\infty$-morphism} if in addition $f_0=0$.
 \label{defn:Linftymorp}
\end{defn}

 \subsection{Open-closed homotopy algebra (OCHA)}
\label{ssec:ocha}

\begin{defn}[Open-closed homotopy algebra (OCHA)\cite{OCHA}]
Let $\cH=\cH_c\oplus\cH_o$ be a graded vector space and 
$(\cH_c,\l)$ be a weak $L_\infty$-algebra. 
Consider a collection of multi-linear maps
\begin{equation*}
 \n:=\{n_{k,l} : (\cH_c)^{\otimes k}\otimes (\cH_o)^{\otimes l}
\raw\cH_o\}_{k,l\ge 0}
\end{equation*}
each of which is graded symmetric on $(\cH_c)^{\otimes l}$. 
We call $(\cH,\l,\n)$ a {\em weak open-closed homotopy algebra (weak OCHA)} 
when $\n$ satisfies the following relations 
\begin{equation}
\begin{split}
&0=\sum_{p+r=n}\sum_{\sigma\in\S_n}
\frac{(-1)^{\epsilon(\sigma)}}{p!\ r!}
n_{1+r,m}(l_p(c_{\sigma(1)},\cdots,c_{\sigma(p)}),
c_{\sigma(p+1)}\cdots,c_{\sigma(n)};o_1,\cdots,o_m) \\
& +
\sum_{p+r=n}\sum_{i+s+j=m}\sum_{\sigma\in\S_n}
\frac{(-1)^{\mu_{p,i}(\sigma)}}{p!\ r!} \\
&\qquad 
n_{p,i+1+j}(c_{\sigma(1)},\cdot\cdot,c_{\sigma(p)};o_1,\cdot\cdot,o_i,
n_{r,s}(c_{\sigma(p+1)},\cdot\cdot,c_{\sigma(n)};o_{i+1},\cdot\cdot,o_{i+s}),
o_{i+s+1},\cdot\cdot,o_m) \ . 
\end{split}
 \label{occd}
\end{equation}
Here the sign exponent $\mu_{p,i}(\sigma)$ is given explicitly by 
\begin{equation}
\mu_{p,i}(\sigma)=\epsilon(\sigma)+
(c_{\sigma(1)}+\cdots +c_{\sigma(p)})+(o_1+\cdots+o_i)+
(o_1+\cdots +o_i)(c_{\sigma(p+1)}+\cdots+c_{\sigma(n)}),
 \label{eta_pi}
\end{equation}
corresponding to the signs effected by the interchanges. 
In particular, if $l_0=n_{0,0}=0$, we call $(\cH,\l,\n)$ an 
{\em open-closed homotopy algebra (OCHA)}. 
We can also write the defining equation (\ref{occd}) 
in the following shorthand expression, 
\begin{equation*}
 \begin{split}
0=& \sum_{p+r=n}\sum_{\sigma\in\S_n}
\frac{(-1)^{\epsilon(\sigma)}}{p!\ r!}
(-1)^{\epsilon(\sigma)}
n_{1+r,m}\left( 
(l_p\otimes \1_c^{\otimes r}\otimes\1_o^{\otimes m})
(c_{\sigma(1)},\cdots,c_{\sigma(n)};o_1,\cdots,o_m) \right) \\
& +\sum_{p+r=n}\sum_{i+s+j=m}\sum_{\sigma\in\S_n}
\frac{(-1)^{\epsilon(\sigma)}}{p!\ r!}
n_{p,i+1+j}\left( (\1_c^{\otimes p}\otimes \1_o^{\otimes i}\otimes
n_{r,s}\otimes \1_o^{\otimes j} )
(c_{\sigma(1)},\cdots,c_{\sigma(n)};o_1,\cdots,o_m) \right)\ ,
 \end{split}
\end{equation*}
where the complicated sign is absorbed into this expression. 
 \label{defn:oc}
\end{defn}
\begin{rem}
For an OCHA $(\cH,\l,\n)$, the substructure 
$(\cH_c,\l)$ is by definition an $L_\infty$-algebra and 
$(\cH_o,\{n_{0,k}\})$ forms an $A_\infty$-algebra. 
Furthermore, 
the substructure $(\cH,\{n_{1,q}\}_{q\ge 0})$ forms an $A_\infty$-module 
over the $A_\infty$-algebra $(\cH_o,\m)$ 
in the sense of \cite{markl:module,Staconf}. 
Also, if $n_{p,0}=0$ for all $p\ge 1$, the substructure
$(\cH,\{n_{p,1}\})$ makes $\cH_o$ an $L_\infty$ -module over $(\cH_c,\l)$ 
\cite{lada-markl}. 
 \label{rem:include}
\end{rem}
Now, let us denote $l_1=d_c$ and $n_{0,1}=d_o$. 
The first few relations which do not appear as 
$A_\infty$ or $L_\infty$ conditions are
\begin{align}
  0=& d_on_{1,0}+n_{1,0}d_c\ , \label{ocrel-10}\\
  0=& d_on_{1,1}(c;o)+n_{1,1}(c;d_o(o))+n_{1,1}(d_c(c);o)\nn\\
    &\quad +n_{0,2}(n_{1,0}(c),o)
 +(-1)^{c(o+1)}n_{0,2}(o,n_{1,0}(c))\ ,\label{ocrel-11}\\
  0=& d_on_{2,0}(c_1,c_2)
 +n_{2,0}(d_c(c_1),c_2)+(-1)^{c_1 c_2}n_{2,0}(d_c(c_2),c_1)\nn\\
  &\quad +n_{1,0}l_2(c_1,c_2) 
   +(-1)^{c_1}n_{1,1}(c_1,n_{1,0}(c_2))
   +(-1)^{c_2(1+c_1)}n_{1,1}(c_2,n_{1,0}(c_1))\ ,\label{ocrel-20} \\
%\end{align}
%\begin{align}
  0=& d_on_{1,2}(c;o_1,o_2)+n_{1,2}(d_c(c);o_1,o_2)
 +(-1)^{c}n_{1,2}(c;d_o(o_1),o_2)+(-1)^{c+o_1}n_{1,2}(c;o_1,d_o(o_2))
 \nn\\ 
 & +n_{1,1}(c,n_{0,2}(o_1,o_2)) 
   +n_{0,2}(n_{1,1}(c;o_1),o_2)
   +(-1)^{o_1(1+c)}n_{0,2}(o_1,n_{1,1}(c,o_2)) \nn\\
 & +n_{0,3}(n_{1,0}(c),o_1,o_2)
   +(-1)^{o_1(1+c)}n_{0,3}(o_1,n_{1,0}(c),o_2)
   +(-1)^{(o_1+o_2)(1+c)}n_{0,3}(o_1,o_2,n_{1,0}(c))\ ,
\label{ocrel-12}
\end{align}
\begin{align}
  0=& d_on_{2,1}(c_1,c_2;o)+n_{2,1}(d_c(c_1),c_2;o)+
 (-1)^{c_1}n_{2,1}(c_1,d_c(c_2);o)
 +(-1)^{c_1+c_2}n_{2,1}(c_1,c_2;d_o(o))\nn \\
 & +n_{1,1}(l_2(c_1,c_2);o)+
   +(-1)^{c_1}n_{1,1}(c_1;n_{1,1}(c_2;o))
  +(-1)^{c_2(c_1+1)}n_{1,1}(c_2;n_{1,1}(c_1;o))\nn\\
 &+n_{0,2}(n_{2,0}(c_1,c_2),o)
 +(-1)^{o(1+c_1+c_2)}n_{0,2}(o,n_{2,0}(c_1,c_2))\nn\\
 &+(-1)^{c_2(1+c_1)}n_{1,2}(c_2;n_{1,0}(c_1),o)
 +(-1)^{(c_2+o)(1+c_1)}n_{1,2}(c_2;o,n_{1,0}(c_1))\nn\\
 &+(-1)^{c_1}n_{1,2}(c_1;n_{1,0}(c_2),o)
 +(-1)^{c_1+o(1+c_2)}n_{1,2}(c_1;o,n_{1,0}(c_2))\ ,
 \label{ocrel-21}\\
 .. & ... \nn
\end{align}
Equation (\ref{ocrel-10}) implies $n_{1,0}$ is a chain map 
by an appropriate relative shift of the grading. 
On the other hand, in the case $n_{0,1}=0$, 
eq.(\ref{ocrel-11}) is an extended Leibniz rule. 
Suppose that we have an OCHA with only non-zero structures 
$d_c,d_o,l_2,n_{1,1},m_2:=n_{0,2}$. 
In eq.(\ref{ocrel-12}) only the second line survives, which 
means that $\cH_c$ acts on an algebra $(\cH_o,m_2)$ by $n_{1,1}$ 
as derivations. 
Furthermore, in eq.(\ref{ocrel-21}) only the second line survives, 
which implies that $\cH_o$ represents a Lie algebra $(\cH_c,l_2)$. 
Then $(\cH,d_c,d_o,l_2,n_{1,1},m_2)$ forms what is called 
a $\g$-algebra or Leibniz pair (see \cite{OCHA} and references there). 
\begin{defn}[Open-closed homotopy algebra (OCHA) morphism]
For two weak OCHAs 
$(\cH,\l,\n)$ and $(\cH',\l',\n')$, 
consider a collection $\f$
of degree zero (degree preserving) multi-linear maps 
\begin{equation*}
 \begin{split}
 f_k &: (\cH_c)^{\otimes k}
 \raw\cH'_c\ ,\qquad \mbox{for}\ {k\ge 0}\ , \\
 f_{k,l} &: (\cH_c)^{\otimes k}\otimes (\cH_o)^{\otimes l}
 \raw\cH'_o\ ,\qquad \mbox{for}\ {k,l\ge 0}\ , 
 \end{split}
\end{equation*}
where $f_k$ and $f_{k,l}$ are graded symmetric with respect to 
$(\cH_c)^{\otimes k}$. 
We call $\f:(\cH,\l,\n)\raw (\cH',\l',\n')$ 
a {\em weak OCHA-morphism} when 
$\{f_k\}_{k\ge 0}:(\cH_c,\l)\raw (\cH_c',\l')$ is a weak 
$L_\infty$-morphism and $\{f_{k,l}\}_{k,l\ge 0}$ further satisfies 
the following relations, 
\begin{equation}
\begin{split}
& \sum_{p+r=n}\sum_{\sigma\in\S_n}
\frac{(-1)^{\epsilon(\sigma)}}{p!\ r!}
f_{1+r,m}\left( 
(l_p\otimes \1_c^{\otimes r}\otimes\1_o^{\otimes m})
(c_{\sigma(1)},\cdots,c_{\sigma(n)};o_1,\cdots,o_m) \right) \\
& +
\sum_{p+r=n}\sum_{i+s+j=m}
\sum_{\sigma\in\S_n}
\frac{(-1)^{\epsilon(\sigma)}}{p!\ r!}
f_{p,i+1+j}\left( (\1_c^{\otimes p}\otimes \1_o^{\otimes i}\otimes
n_{r,s}\otimes \1_o^{\otimes j} )
(c_{\sigma(1)},\cdots,c_{\sigma(n)};o_1,\cdots,o_m) \right) \\
&=
\sum_{\substack{(r_1+\cdots +r_i)+(p_1+\cdots +p_j)=n\\
(q_1+\cdots +q_j)=m}}
\sum_{\sigma\in\S_n}
\frac{(-1)^{\epsilon(\sigma)}}{i! (r_1!\cdots r_i!) (p_1!\cdots p_j!)}\\
&\hspace*{4.0cm}
n'_{i,j}\left( (f_{r_1}\otimes\cdot\cdot\otimes 
f_{r_i}\otimes
f_{p_1, q_1}\otimes\cdot\cdot\otimes f_{p_j,q_j})
(c_{\sigma(1)},\cdots,c_{\sigma(n)};o_1,\cdots,o_m) \right) \ . 
\end{split}
 \label{ocmorpcd}
\end{equation}
The right hand side is written explicitly as 
\begin{equation*}
 \begin{split}
& n'_{i,j}\left( (f_{r_1}\otimes\cdots\otimes 
f_{r_i}\otimes
f_{p_1, q_1}\otimes\cdots\otimes f_{p_j,q_j})
(c_{\sigma(1)},\cdots,c_{\sigma(n)};o_1,\cdots,o_m) \right) \\
&=(-1)^{\tau_{\oraw{p},\oraw{q}}(\sigma)}
n'_{i,j}(f_{r_1}
(c_{\sigma(1)},\cdot\cdot,c_{\sigma(r_1)}),
\cdots\cdot,f_{r_i}
(c_{\sigma({\bar r}_{i-1}+1)},\cdot\cdot,c_{\sigma({\bar r}_i)}); \\
&\hspace*{1.0cm}
f_{p_1, q_1}
(c_{\sigma({\bar r}_i+1)},\cdot\cdot,c_{\sigma({\bar p}_1)}
;o_1,\cdot\cdot,o_{q_1}),
\cdots\cdot,f_{p_j,q_j}
(c_{\sigma({\bar p}_{j-1}+1)},\cdot\cdot,c_{\sigma({\bar p}_j)};
o_{{\bar q}_{j-1}+1},\cdot\cdot,o_{{\bar q}_j})
)\ , 
 \end{split}
\end{equation*}
where 
${\bar r}_k:=r_1+\cdots +r_k$, 
${\bar p}_k:={\bar r}_i+ p_1+\cdots +p_k$, 
${\bar q}_k:=q_1+\cdots +q_k$ 
and 
$\tau_{\oraw{p},\oraw{q}}(\sigma)$ is given by 
$$
\tau_{\oraw{p},\oraw{q}}(\sigma)=
\sum_{k=1}^{j-1}
(c_{\sigma({\bar p}_k+1)}+\cdots +c_{\sigma({\bar p}_{k+1})})
(o_1+\cdots +o_{{\bar q}_k}) \ .
$$
In particular, if $(\cH,\l,\n)$ and $(\cH',\l',\n')$ are 
OCHAs and if $f_0=f_{0,0}=0$, 
we call it an {\em OCHA-morphism}. 
 \label{defn:ocmorp}
\end{defn}
\begin{defn}[Quasi-isomorphism]
Suppose that two OCHAs $(\cH,\l,\n)$, 
$(\cH',\l',\n')$ and an OCHA-morphism 
$\f: (\cH,\l,\n)\raw (\cH',\l',\n')$ are given. 
$\f$ is called an {\it open-closed homotopy algebra quasi-isomorphism} 
if $f_1:\cH_c\raw\cH'_c$ induces an isomorphism 
between the cohomology spaces of the complexes $(\cH_c,l_1)$ and 
$(\cH'_c,l_1')$, and further 
$f_{0,1}:\cH_o\raw\cH'_o$ induces an isomorphism 
between the cohomology spaces of 
the complexes $(\cH_o,n_{0,1})$ and $(\cH'_o,n'_{0,1})$. 
In particular, if $f_1$ and $f_{0,1}$ are isomorphisms, 
we call $\f$ an open-closed homotopy algebra isomorphism. 
 \label{defn:quasiisom}
\end{defn}

 \subsection{Cyclic structures in OCHAs}
\label{ssec:cyclic}

Now we consider an additional structure, cyclicity, (cf. \cite{GeK1}) 
on open-closed homotopy algebras. 
It is defined 
in terms of constant symplectic inner products.
The string theory motivation for this additional structure 
is that punctures on the boundary
of the disk inherit a cyclic order from the orientation of the disk
and the operations are to respect this cyclic structure, 
just as the $L_\infty$-structure reflects the symmetry 
of the punctures in the interior of the disk or on the sphere. 
Alternatively, a typical Lagrangian of a (quantum) field theory originally 
has such structure and in particular in the 
Batalin-Vilkovisky (BV) formalism \cite{BV1,BV2}, 
such structure is defined in terms of an odd (degree minus one) 
symplectic structure on the corresponding supermanifold (\ref{sec:dual}
\cite{Sch,AKSZ,thesis}. 
Both pictures are then combined with each other in 
string field theory as discussed in section \ref{sec:ocsft}. 

{}From such background, in \cite{thesis} 
a ``cyclicity" is defined for $A_\infty$-algebras 
in terms of a degree minus one constant symplectic inner product, 
and it is shown that homotopy invariant properties of
$A_\infty$-algebras hold true also in the category of cyclic 
$A_\infty$-algebras. 
However, in string theory or in particular topological string theory, 
there often exist cyclic structures defined by inner products having some
different degree. 
For the arguments on  homotopy invariant properties 
in \cite{thesis}, the degree of the inner product is not essential. 
Thus, we define cyclic structures with constant symplectic 
inner products of arbitrary fixed integer degrees. 
\begin{defn}[Constant symplectic structure]
Bilinear maps, 
$\omega_c: \cH_c\otimes\cH_c\raw\C$ and 
$\omega_o: \cH_o\otimes\cH_o\raw\C$, 
are called {\em constant symplectic structures} 
when they have fixed integer degrees 
$|\omega_c|, |\omega_o|\in\Z$ and are non-degenerate and skew-symmetric. 
Here `skew-symmetric' indicates that 
\begin{equation*}
 \omega_c(c_2,c_1)=-(-1)^{c_1c_2}\omega_c(c_1,c_2)\ ,\qquad 
 \omega_o(o_2,o_1)=-(-1)^{o_1o_2}\omega_o(o_1,o_2)\ 
\end{equation*} 
for any $c_1,c_2\in\cH_c$, $o_1,o_2\in\cH_o$, and 
the degree of $\omega_c,\omega_o$ implies that 
$\omega_c(c_1,c_2)=0$ except for $\deg(c_1)+\deg(c_2)+|\omega_c|=0$ and 
$\omega_o(o_1,o_2)=0$ except for $\deg(o_1)+\deg(o_2)+|\omega_o|=0$. 
We further denote the constant symplectic structure on 
$\cH=\cH_c\oplus\cH_o$ by $\omega:=\omega_c\oplus\omega_o$. 
 \label{defn:sym}
\end{defn}
Suppose that an open-closed homotopy algebra $(\cH,\l,\n)$ 
is equipped with constant symplectic structures 
$\omega_c: \cH_c\otimes\cH_c\raw\C$ and 
$\omega_o: \cH_o\otimes\cH_o\raw\C$ as in Definition \ref{defn:sym}.

For $\{l_k\}_{k\ge 1}$ and $\{n_{p,q}\}_{p+q\ge 1}$, 
let us define two kinds of multi-linear maps by 
\begin{equation*}
 \V_{k+1}=\omega_c(l_k\ott\1_c): (\cH_c)^{\otimes (k+1)}\raw\C\
 ,\qquad 
 \V_{p,q+1}=\omega_o(n_{p,q}\ott\1_o):(\cH_c)^{\otimes p}\otimes 
(\cH_o)^{\otimes (q+1)}\raw\C\ 
\end{equation*}
or more explicitly 
\begin{equation*}
 \V_{k+1}(c_1,\cdots,c_{k+1})
=\omega_c(l_k(c_1,\cdots,c_k),c_{k+1}) 
\end{equation*}
and 
\begin{equation*}
 \V_{p,q+1}(c_1,\cdots,c_p;o_1,\cdots,o_{q+1})=
 \omega_o(n_{p,q}(c_1,\cdots,c_p;o_1,\cdots,o_{q}),o_{q+1})\ .
\end{equation*}
The degree of $\V_{k+1}$ and $\V_{p,q+1}$ are 
$|\omega_c|+1$ and $|\omega_o|+1$. 
Note that the degrees of $\V_{k+1}$ and $\V_{p,q+1}$ are zero 
when they come from odd constant symplectic structures 
$|\omega_c|=|\omega_o|=-1$. 
\begin{defn}[Cyclic open-closed homotopy algebra (COCHA)]
An open-closed homotopy algebra $(\cH,\omega,\l,\n)$ 
is called a {\em cyclic open-closed homotopy algebra} (COCHA)
when 
$\V_{k+1}$ is graded symmetric with respect to 
any permutation of $(\cH_c)^{\otimes(k+1)}$ and 
$\V_{p,q+1}$ has cyclic symmetry 
with respect to cyclic permutations of $(\cH_o)^{\otimes (q+1)}$, 
that is, if 
\begin{equation*}
 \V_{k+1}(c_1,\cdots,c_{k+1})
 =(-1)^{\epsilon(\sigma)}\V_{k+1}(c_{\sigma(1)},\cdots,c_{\sigma(k+1)})\ , 
 \qquad \sigma\in\S_{k+1}\ 
\end{equation*}
and 
\begin{equation*}
 \V_{p,q+1}(c_1,\cdots,c_p;o_1,\cdots,o_{q+1})=
(-1)^{o_1(o_2+\cdots +o_{q+1})}
\V_{k+1,l}(c_1,\cdots,c_p;o_2,\cdots,o_{q+1},o_1)\ . 
\end{equation*} 
The graded commutativity of $\V_{p,q+1}$ with respect to permutations of
$(\cH_c)^{\ott p}$, that is, 
\begin{equation*}
 \V_{p,q+1}(c_1,\cdots,c_p;o_1,\cdots,o_{q+1})=
 (-1)^{\epsilon(\sigma)}\V_{p,q+1}
(c_{\sigma(1)},\cdots,c_{\sigma(p)};o_1,\cdots,o_{q+1})\ ,
 \qquad \sigma\in\S_p
\end{equation*}
automatically holds by the definition of $\n$. 
 \label{defn:cyc} 
\end{defn}
Note also that there are many situations where the inner products 
exist only for open strings. 
This is  the case for the topological string situation 
in the B-model we will discuss later 
in subsection \ref{ssec:Bmodel}. 
For instance, on the topological open string side, there often 
exists a natural inner product coming essentially from 
an integral (trace) of products of two differential forms. 
The inner products of this kind in fact turn out to be 
skew-symmetric (symplectic) 
in our suspended notation (see \cite{fol2}). 
See also \cite{Poli} for more general cyclic structures including 
`non-skew' inner products.

On the other hand, if 
we have $\omega_c$ and $\omega_o$, 
non-degenerate inner products in 
both open and closed string sides, 
we can identify $\cH$ with its linear dual, 
then reverse the process and define further maps
$$
 r_{p-1,q+1}: (\cH_c)^{\otimes (p-1)}\otimes (\cH_o)^{\otimes (q+1)}
 \to \cH_c
$$
with relations amongst themselves and with the operations already 
defined, which can easily be deduced from their definition.
In particular, for $n_{1,0}:\cH_c\raw\cH_o$ we have 
$r_{0,1}: \cH_o\raw\cH_c$.  
Namely, for the cyclic case 
the fundamental object is the multi-linear map $\V_{p,q+1}$ 
where $n_{p,q}$ and $r_{p-1,q+1}$ are equivalent under the 
relation above. 

Physically, the multi-linear map $\V_{p,q+1}$ is related to 
the (scattering) amplitudes of a disk with 
$p$ closed strings and $(q+1)$ open strings insertions. 
Choosing an open string state as a root edge instead of a closed
string state, that is, taking $n_{p,q}$ instead of $r_{p-1,q+1}$, 
for defining an OCHA 
is related to a standard compactification of the moduli spaces of 
the corresponding Riemann surface 
(a disk with $p$ points interior and $(q+1)$ points on the boundary). 
Also, in the next section we shall see that, 
due to this choice of the root edge, the OCHA structure $(\l,\n)$ 
can be singled out to be an odd vector field 
on the appropriate supermanifold.

\begin{rem}[Category version]
As an $A_\infty$-category is defined as a straightforward extension of 
an $A_\infty$-algebra \cite{Fukaya}, one can extend our open-closed
homotopy algebra 
to its category version by replacing $\cH_o$ by the space of 
morphisms of a category.
This category extension corresponds to considering many D-branes 
on which open strings end. 
This is important for applying OCHAs to topological string theory, 
see subsection \ref{ssec:Bmodel}. 
\end{rem}

\section{The dual supermanifold description} 
\label{sec:dual}

 \subsection{OCHAs and odd formal vector fields}
\label{ssec:vectfd}

For a graded vector space $\cH=\cH_c\oplus\cH_o$, 
denote by $\{\eb_{c,i}\}$ a basis of $\cH_c$ and 
by $\{\eb_{o,i}\}$ a basis of $\cH_o$. 
For each $\eb_{c,i}\in\cH_c$ represent the 
dual base as $\psi^i$ and similarly 
the dual base of $\eb_{o,i}\in\cH_o$ as $\phi^i$. 
We set the degree of the dual basis by 
$\deg(\psi^i)=-\deg(\eb_{c,i})$ and $\deg(\phi^i)=-\deg(\eb_{o,i})$. 
We consider the formal power series ring in the variables 
$\{\phi^i\}$, $\{\psi^i\}$ and $\{\phi^i\}\coprod\{\psi^i\}$, 
and denote them by $C(\phi)$, $C(\psi)$ and $C(\psi,\phi)$, respectively. 
We define $\{\psi^i\}$ to be associative and graded commutative 
and $\{\phi^i\}$ to be associative but noncommutative. 
More precisely, 
in the space of the formal power series 
of associative fields $\{\phi^i\}\coprod\{\psi^i\}$, 
an element in $\{\psi^i\}$ is graded commutative with respect to all elements. 
Therefore, any element in $C(\psi,\phi)$ is represented as 
\begin{equation}
  a(\psi,\phi)=\sum_{k,l}a_{i_1\cdots i_k;j_1\cdots j_l}
 (\phi^{j_l}\cdots\phi^{j_1})(\psi^{i_k}\cdots\psi^{i_1})\ ,
 \label{a}
\end{equation}
where the coefficient $a_{i_1\cdots i_k;j_1\cdots j_l}\in\C$ is 
graded symmetric with respect to $i_1\cdots i_k$. 
We can call $(\cH,C(\phi,\psi))$ 
the {\em  formal supermanifold} \cite{AKSZ,Ko1} 
corresponding to an OCHA $(\cH,\l,\n)$. 
Though usually the term `super' indicates $\Z_2$-graded, 
we use it for $\Z$-graded objects. 
On the other hand, an $A_\infty$-algebra is described on a formal 
noncommutative supermanifold $(\cH_o,C(\phi))$ \cite{Ka,thesis}, and 
an $L_\infty$-algebra is described on 
a formal graded commutative supermanifold $(\cH_c,C(\psi))$. 

For a weak OCHA $(\cH,\l,\n)$, 
express the collection of multi-linear maps
\begin{equation*}
 l_k : (\cH_c)^{\otimes k}\raw\cH_c\ ,\qquad 
 n_{k,l} : (\cH_c)^{\otimes k}\otimes (\cH_o)^{\otimes l}\raw\cH_o\ ,
\end{equation*}
in terms of the bases: 
\begin{equation*}
 \begin{split}
 &l_k(\eb_{c,i_1},\cdots,\eb_{c,i_k})
 =\eb_{c,j}c^j_{i_1\cdots i_k}\ ,\qquad 
 c^j_{i_1\cdots i_k}\in\C\ ,\\
 &n_{k,l}(\eb_{c,i_1},\cdots,\eb_{c,i_k};\eb_{o,j_1},\cdots,\eb_{o,j_l})
 =\eb_{o,j}c^j_{i_1\cdots i_k;j_1\cdots j_l}\ ,\qquad 
c^j_{i_1\cdots i_k;j_1\cdots j_l}\in\C\ .
 \end{split}
\end{equation*}
Correspondingly, let us define an odd formal vector
field $\delta:=\delta_S+\delta_D : C(\phi,\psi)\raw C(\phi,\psi)$, where 
\begin{equation}
\begin{split}
 \delta_S
 &=\flpartial{\psi^j}c^j(\psi)
 =\sum_{k\ge 0}\ov{k!}
 \flpartial{\psi^j}c^j_{i_1\cdots i_k}
 \psi^{i_k}\cdots\psi^{i_1}\ ,\qquad c^j(\psi)\in C(\psi)\ , \\
 \delta_D
 &=\flpartial{\phi^j}c^j(\phi,\psi)
 =\sum_{k+l\ge 0}\ov{k!}
 \flpartial{\phi^j}c^j_{i_1\cdots i_k;j_1\cdots j_l}
 (\phi^{j_l}\cdots\phi^{j_1})(\psi^{i_k}\cdots\psi^{i_1})\ ,
 \qquad c^j(\phi,\psi)\in C(\phi,\psi)\ .
 \end{split}
 \label{delta}
\end{equation}
We use right derivatives just for the sign problem; 
it is easy to relate this dual supermanifold description 
to the convention in the previous section. 
Since $\l$ and $\n$ have degree one, $\delta$ also has degree one. 

It acts on $C(\phi,\psi)$ as follows:
\begin{equation}
 \begin{split}
 &\delta\left(a(\phi,\psi)\right)
 =\sum_{k,l}a_{i_1\cdots i_k;j_1\cdots j_l}
 \sum_{s=1}^k(-1)^{\beta_S(s-1)}
 (\phi^{j_l}\cdots\phi^{j_1})
 (\psi^{i_k}\cdot\cdot\delta_S(\psi^{i_s})\cdot\cdot\psi^{i_1})\\
 &\qquad +\sum_{k,l}a_{i_1\cdots i_k;j_1\cdots j_l}
 \sum_{t=1}^l(-1)^{\beta_D(t-1)}
 (\phi^{j_l}\cdot\cdot\delta_D(\phi^{j_t})\cdot\cdot\phi^{j_1})
 (\psi^{i_k}\cdots\psi^{i_1})\ ,
 \end{split}\label{deltaact}
\end{equation}
where $\delta_D(\phi^{j_t})=c^{j_t}(\phi,\psi)$ and 
$\delta_S(\psi^{i_s})=c^{i_s}(\psi)$. 
Then, $\beta_S(s-1)$ (resp. $\beta_D(t-1)$) is the sign arising when 
$\delta_S$ (resp. $\delta_D$) acts from the right and passes the 
corresponding superfields and is given explicitly by 
\begin{equation*}
 \beta_S(s-1)=\eb_{c,i_1}+\cdots +\eb_{c,i_{s-1}}\ ,\qquad 
 \beta_D(t-1)=(\eb_{c,i_1}+\cdots +\eb_{c,i_k})
 +\eb_{o,j_1}+\cdots +\eb_{o,j_{t-1}}\ .
\end{equation*}
The above $\delta\left(a(\phi,\psi)\right)$ is further rewritten; 
in the first line 
$\psi$'s in $\delta_D(\phi^{j_t})$ are brought to the right of $\phi$'s, 
and $\psi$'s on each line of eq.(\ref{deltaact}) 
are treated as graded symmetric.  
The $\delta\left(a(\phi,\psi)\right)$ is expressed 
in the form in eq.(\ref{a}) again 
but with a different coefficient. 
In the supermanifold language,
$\delta$ is called an 
(odd) {\it formal vector field} on the formal supermanifold. 
A formal manifold with such a $\delta$ is called 
a {\it $Q$-manifold} in \cite{AKSZ}
if $Q = \delta$ with $Q^2=0$. 
\begin{lem}
The condition that $(\cH,\l,\n)$ is a weak OCHA 
is equivalent to 
\begin{equation}
 (\delta)^2=0\ .
\end{equation}
In particular,  $\delta$ defines an OCHA 
if the $k=0$ part of $\delta_S$ and $k=l=0$ part of $\delta_D$ in 
eq.(\ref{delta}) are absent or zero. 
 \label{lem:dualoccd}
\end{lem}
The equation above can be expanded as 
$(\delta_S)^2+[\delta_S,\delta_D]+(\delta_D)^2=0$, 
where $[\delta_S,\delta_D]=\delta_S(\delta_D)+\delta_D(\delta_S)$, 
$\delta_S(\delta_D)=\flpartial{\phi^j}(\delta_S(c^j(\phi,\psi))).$ 
Note that $\delta_D(\delta_S)$ vanishes 
since $\delta_S$ does not include $\phi$. 
Furthermore, one can see that 
\begin{equation*}
 (\delta_S)^2=0\ ,\qquad \delta_S(\delta_D)+(\delta_D)^2=0\  
\end{equation*}
hold independently. 
The first one is just the dual of the 
$L_\infty$-condition (\ref{Linfty}), 
and the second one is the dual description of 
the OCHA condition (\ref{occd}). 
The pair of the equations above can also be thought of as 
a deformation of $\delta_S$ by $\delta_S + \delta_D$, 
though we do not discuss this type of deformation in this paper. 

A (weak) OCHA morphism 
in Definition \ref{defn:ocmorp} can also be rewritten in the same way. 
For the collection $\f$ of degree zero multi-linear maps, 
\begin{equation*}
 f_l : (\cH_c)^{\otimes l}\raw\cH'_c\ ,\qquad 
 f_{k,l} : (\cH_c)^{\otimes k}\otimes (\cH_o)^{\otimes l}
 \raw\cH'_o\ ,
\end{equation*}
let us now express $f_k$ and $f_{k,l}$  as 
\begin{equation}
 \begin{split}
 &f_k(\eb_{c,i_1},\cdots,\eb_{c,i_k})
=\eb_{c,j'}f^{j'}_{i_1\cdots i_k}\ ,\qquad 
 f^{j'}_{i_1\cdots i_k}\in\C\ ,\\
 &f_{k,l}(\eb_{c,i_1},\cdots,\eb_{c,i_k};\eb_{o,j_1},\cdots,\eb_{o,j_l})
 =\eb_{o,j'}f^{j'}_{i_1\cdots i_k;j_1\cdots j_l}\ ,\qquad 
 f^{j'}_{i_1\cdots i_k;j_1\cdots j_l}\in\C\ .
 \end{split}
\end{equation}
They define the following coordinate transformation between the two 
supermanifolds $(\cH,\l,\n)$ and $(\cH',\l',\n')$, 
\begin{equation}
 \begin{split}
 \psi^{j'}&=\f_*^{j'}(\psi)
 =f^{j'}+f^{j'}_i\psi^i+f^{j'}_{i_1i_2}\psi^{i_2}\psi^{i_1}
 +\cdots+f^{j'}_{i_1\cdots i_n}\psi^{i_n}\cdots\psi^{i_1}+\cdots\ ,\\
 \phi^{j'}&=\f_*^{j'}(\phi,\psi)
 =\sum_{k,l\ge 0}f^{j'}_{i_1\cdots i_k;j_1\cdots j_l}
(\phi^{j_l}\cdots\phi^{j_1})(\psi^{i_k}\cdots\psi^{i_1})\ .
 \end{split}
 \label{coordtransf}
\end{equation}
This induces a pullback from $C(\phi',\psi')$ to $C(\phi,\psi)$, 
\begin{equation*}
 \f^*\left(a(\phi',\psi')\right)=a(\f_*(\phi,\psi),\f_*(\psi))\ ,
\end{equation*} 
where $\{\phi^i\}$ and $\{\phi^{i'}\}$ are the coordinates 
on $\cH$ and $\cH'$, respectively. 
\begin{lem}
The condition that this $\f$ is a weak 
OCHA morphism is then 
that the map between two formal supermanifolds $\f_*$ 
is compatible with the actions of $\delta$ and $\delta'$ on both
sides, that is, 
\begin{equation}
 \f^*\delta'(a(\phi',\psi'))=\delta\f^* a(\phi',\psi')
\end{equation}
holds for any $a(\phi',\psi')\in C(\phi',\psi')$. 
In other words, $\f_*$ is a morphism between $Q$-manifolds. 
If in addition $f^{j'}=f^{j'}_{\emptyset;\emptyset}=0$, 
$\f_*$ preserves the origin of
the formal supermanifolds. $\f_*$ is then 
an OCHA morphism in the situation that 
both $\delta$ and $\delta'$ define  OCHAs. 
 \label{lem:dualocmorp}
\end{lem}
All these structures in the supermanifold description 
are dual to the coalgebra description explained in \cite{OCHA} 
in the following sense 
(see \cite{thesis} for the $A_\infty$ case). 
Let us introduce natural pairings 
\begin{equation*}
 \la \psi^i|\eb_{c,j}\ra=\delta^i_j\ ,\qquad 
 \la \phi^i|\eb_{o,j}\ra=\delta^i_j\ 
\end{equation*}
and also the extended pairings 
\begin{equation*}
 \la (\phi^{j_l}\cdots\phi^{j_1})(\psi^{i_k}\cdots\psi^{i_1})
 |(\eb_{c,i'_1}\cdots\eb_{c,i'_k})\otimes
(\eb_{o,j'_1}\cdots\eb_{o,j'_l}) \ra
 =\epsilon^{i_1\cdots i_k}_{i'_1\cdots i'_k}
 \delta^{j_1}_{j'_1}\cdots\delta^{j_k}_{j'_k}\ 
\end{equation*}
for $k+l>1$, where 
$\epsilon^{i_1\cdots i_k}_{i'_1\cdots i'_k}:
=\sum_{\sigma\in\S_k}\epsilon(\sigma)
\delta^{i_{\sigma(1)}}_{i'_1}\cdots\delta^{i_{\sigma(k)}}_{i'_k}$. 
We set the pairing to be zero if the number of elements 
$\psi/\phi$ and $\eb_c/\eb_o$ does not coincide. 
The space spanned by 
$(\eb_{c,i_1}\cdots\eb_{c,i_k})\otimes(\eb_{o,j_1}\cdots\eb_{o,j_l})$, 
$k+l\ge 1$, is what is denoted $C(\cH_c)\otimes T^c(\cH_o)$ 
in \cite{OCHA}. 
As the adjoint of the product on $C(\psi,\phi)$, 
a coproduct 
$\Delta:C(\cH_c)\otimes T^c(\cH_o)\to
(C(\cH_c)\otimes T^c(\cH_o))\otimes(C(\cH_c)\otimes T^c(\cH_o))$ 
is defined; 
for $a,b\in C(\psi,\phi)$ and $x\in C(\cH_c)\otimes T^c(\cH_o)$, 
\begin{equation*}
 \la b\cdot a| x\ra = 
 \sum_i \la a | x_i^+\ra\cdot\la b|x_i^-\ra\ ,\qquad 
 \Delta:=\sum_i(x_i^+\otimes x_i^-)\ .
\end{equation*}
Thus, $C(\cH_c)\otimes T^c(\cH_o)$ forms a coalgebra. 
Then, a codifferential 
$\l+\n: C(\cH_c)\otimes T^c(\cH_o)\to C(\cH_c)\otimes T^c(\cH_o)$ 
is given as the adjoint of $\delta=\delta_S+\delta_D$: 
\begin{equation*}
 \la a |(\l+\n)(x) \ra := \la \delta(a)| x\ra\ .
\end{equation*}
In a similar way, a coalgebra homomorphism 
$\f:C(\cH_c)\otimes T^c(\cH_o)\to C(\cH'_c)\otimes T^c(\cH'_o)$ 
is obtained as the adjoint of the pullback 
$\f^*: C(\phi',\psi')\to C(\phi,\psi)$.  
Thus, the coalgebra description given in \cite{OCHA} is 
obtained as the dual of the algebra description 
in terms of formal power series on supermanifolds.

 \subsection{Cyclicity and constant symplectic/Poisson structures}
\label{ssec:dualcyclic}

Next, we discuss the cyclicity (Definition \ref{defn:cyc}). 
If cyclicity is imposed on the  $\phi^i$'s, we indicate that by  
$C(\phi)_{cyc}$ or $C(\phi,\psi)_{cyc}$. 
Any element of $C(\phi,\psi)_{cyc}\subset C(\phi,\psi)$ is 
represented in the form in eq.(\ref{a}) but the coefficient 
$a_{i_1\cdots i_k;j_1\cdots j_l}$ is in addition graded cyclic
symmetric with respect to the indices $j_1\cdots j_l$. 
On this algebra, a constant Poisson structure is introduced 
naturally by dualizing the constant symplectic structures 
in Definition \ref{defn:sym}. 
\begin{defn}[Constant Poisson structure]
Suppose $\cH_c$ and $\cH_o$ have constant symplectic structures 
$\omega_c: \cH_c\otimes\cH_c\raw\C$ and 
$\omega_o: \cH_o\otimes\cH_o\raw\C$ 
(Definition \ref{defn:sym}). 
The corresponding Poisson brackets are denoted by 
\begin{equation*}
 (\ ,\ )_c=\flpartial{\phi^i}\omega_c^{ij}\frpartial{\phi^j}\ ,
 \qquad 
 (\ ,\ )_o=\flpartial{\phi^i}\omega_o^{ij}\frpartial{\phi^j}\ .
\end{equation*}
Here $\omega_c^{ij}\in\C$ and $\omega_o^{ij}\in\C$ are the inverse
matrices of $\omega_{c,ij}:=\omega_c(\eb_{c,i},\eb_{c,j})$ 
and $\omega_{o,ij}:=\omega_o(\eb_{o,i},\eb_{o,j})$. That is, 
$\omega_{c,ij}\omega_c^{jk}=\omega_c^{kj}\omega_{c,ji}=\delta_i^k$ and 
$\omega_{o,ij}\omega_o^{jk}=\omega_o^{kj}\omega_{o,ji}=\delta_i^k$ hold. 
Thus $(\ ,\ )_c$ is a graded Poisson bracket for 
a graded commutative algebra and 
$(\ ,\ )_o$ is a Poisson bracket for the cyclic algebra as 
in \cite{Ka}. 
$C(\psi)_{cyc}$ and $C(\phi)_{cyc}$ form graded Poisson algebras 
with Poisson brackets $(\ ,\ )_c$ and $(\ ,\ )_o$, respectively. 
Furthermore, these two Poisson brackets can be combined 
naturally and extended to one on $C(\phi,\psi)_{cyc}$. 
 \label{defn:Poisson}
\end{defn}
A COCHA (Definition \ref{defn:cyc}) 
is dualized as follows. 
For the collection of multi-linear maps $\V_k$ and $\V_{k,l}$, 
let us define their coefficients by 
\begin{equation*}
\V_k(\eb_{c,i_1},\cdots,\eb_{c,i_k})
 :=\V_{i_1\cdots i_k}\in\C\ ,\qquad 
\V_{k,l}(\eb_{c,i_1},\cdots,\eb_{c,i_k};\eb_{o,j_1},\cdots,\eb_{o,j_l})
 :=\V_{i_1\cdots i_k; j_1\cdots j_l}\in\C\ .
\end{equation*}
Note that they are graded symmetric 
with respect to the indices $i_1\cdots i_k$ and cyclic 
with respect to the indices $j_1\cdots j_l$. 
Consider further a formal sum of polynomial functions $S$,  
\begin{equation}
 S(\phi,\psi)=S_S(\psi)+ S_D(\phi,\psi)\ ,
 \qquad S_S(\psi)\in C(\psi),\ S_D(\phi,\psi)\in C(\phi,\psi)_{cyc}\ ,
 \label{action1}
\end{equation}
where $S_S$ and $S_D$ are defined by 
\begin{equation}
 \begin{split}
 S_S(\psi)&=\sum_{l\ge 2}\ov{l!}\V_{i_1\cdots i_l}\psi^{i_l}\cdots\psi^{i_1}
 \ ,\qquad \V_{i_1\cdots i_l}\in\C \\
 S_D(\phi,\psi)
 &= \sum_{k\ge 0,l\ge 1, k+l\ge 2}
 \ov{k!\ l}\V_{i_1\cdots i_k;j_1\cdots j_l}(\phi^{j_l}\cdots\phi^{j_1})
 (\psi^{i_k}\cdots\psi^{i_1})\ ,\qquad \V_{i_1\cdots i_k;j_1\cdots
 j_l}\in\C. 
 \end{split}\ 
 \label{action2}
\end{equation}
Then one can define the formal vector field $\delta$ acting on 
$C(\phi,\psi)_{cyc}$ as follows: 
\begin{equation}
 \delta:=\delta_S+\delta_D\ ,\qquad 
 \delta_S=(\ ,S_S)_c\ ,\quad \delta_D=(\ ,S_D)_o\ .
 \label{fvf}
\end{equation}
The condition $(\delta)^2=0$ coincides 
with the condition that $(\cH,\l,\n)$ is a COCHA.

\section{Zwiebach's open-closed string field theory}
\label{sec:ocsft}

String field theory is defined on a fixed conformal background 
of a space-time (target space) $M$ to which world sheets of strings 
(Riemann surfaces) are mapped, 
where the conformal background is a background (metric, etc. of $M$) 
in which string world sheet theory has conformal symmetry. 
There exist several classes of string field theories corresponding 
to the classes of Riemann surfaces. 
The most general one is quantum open-closed string field theory \cite{Z2}, 
which is associated to the most general class of Riemann surfaces: 
Riemann surfaces of arbitrary genus, possibly with boundaries and punctures. 

It includes various `sub-string field theories' : 
classical open string field theories - associated to 
disks (one boundary and zero genus) with punctures only on the boundary, 
classical closed string field theories - associated to 
spheres (no boundary and genus zero) with punctures, 
quantum closed string field theories - 
associated to Riemann surfaces with punctures (and various genera) and 
without boundary, and so on. 
Genera and multi-boundaries relate to closed and open string loops
(in the sense of Feynman diagrams), 
respectively. 
We use the term `classical' (resp. quantum) 
for theory without loops (resp. with loops). 
In this section we shall explain that, extracting the tree open-closed
part from  Zwiebach's quantum open-closed string
field theory \cite{Z2}, one obtains an OCHA. 
Namely, an OCHA is a general homotopy algebraic structure for 
tree open-closed string field theory as $L_\infty$-algebras 
(resp. $A_\infty$-algebras) are 
for tree closed (resp. open) string field theories.

The quantum open-closed string field theory 
discussed by B.~Zwiebach \cite{Z2} 
is defined by all possible open-closed interaction vertices 
together with closed and open string kinetic terms 
satisfying the quantum BV-master equation. 
The interaction term is expressed formally in the following form 
(eq.(5.7) of \cite{Z2}), 
\def\bra#1{\langle #1 |}
\def\ket#1{| #1 \rangle}
\def\cMb{{\bar {\cal M}}}
\begin{equation}
f (\V^{g,n}_{b,m}) 
=\Bigl[  {1\over n!} {1\over b!} \prod_{k=1}^b {1\over m_k} \Bigr] \, 
\int_{\V^{g,n}_{b,n}}\bra{\Omega}
\underbrace{\ket{\Psi}\cdots\ket{\Psi} }_{n} 
\prod_{k=1}^b  \underbrace{\ket{\Phi} \cdots  \ket{\Phi} }_{m_k}\ .
 \label{eq5.7}
\end{equation}
Here the kets $\ket{\Psi}\in \cH_c$ and $\ket{\Phi}\in \cH_o$ 
are the {\em closed string fields} 
and the {\em open string fields}, respectively. 
$\V^{g,n}_{b,m}\subset \cMb^{g,n}_{b,m}$ 
is the appropriate subspace of the 
compactified moduli space of Riemann surfaces with genus
$g$, $n$-interior punctures and $b$ boundaries $S^1$ 
having $m_1,\cdots,m_b$ punctures on them. 
Equivalently, 
it has $n$ closed string punctures 
and $m_i$ open string insertions 
on the corresponding boundary $S^1$. 
The bra $\bra{\Omega}$ denotes a differential form on $\cMb^{g,n}_{b,m}$ 
which takes its value in 
$(\cH_c^*)^{\otimes n}\otimes (\cH_o^*)^{\otimes m_1}\otimes\cdots\otimes
(\cH_o^*)^{\otimes m_b}$. 
This data is determined by the conformal field theory for 
a fixed conformal background. 
Then, the combination $\int_{\V^{g,n}_{b,m}}
\bra{\Omega}$ defines a map
\begin{equation}
\int_{\V^{g,n}_{b,m}}
\bra{\Omega}:(\cH_c)^{\otimes n}\otimes (\cH_o)^{\otimes m_1}
\otimes\cdots\otimes (\cH_o)^{\otimes m_b}\raw \C\ .
 \label{mapOmega}
\end{equation}
In terms of bases $\eb_{c,i}$ and $\eb_{o,j}$ of
the {\em spaces of states} $\cH_c$ and $\cH_o$, 
the kets can be expressed as 
$\ket{\Psi}:=\sum_i\eb_{c,i}\psi^i$ and 
$\ket{\Phi}:=\sum_j\eb_{o,j}\phi^j$, 
and the coordinates $\psi^i$ and $\phi^i$ play the role of fields. 
The degree of each basis element $\eb_{c,i}$ or $\eb_{o,j}$ 
is determined by the 
corresponding  conformal field theory on the string world sheet
and is related to the degree of field $\psi^i$ or 
$\phi^j$ through the relations 
$\deg(\psi^i)=-\deg(\eb_{c,i})$ and $\deg(\phi^i)=-\deg(\eb_{o,i})$. 
The degrees $\deg(\psi^i)$ and $\deg(\phi^i)$ in turn denote the 
ghost numbers in the sense of the BV-formalism for 
the target space field theory. 
The map (\ref{mapOmega}) 
is defined to be of degree zero because of a ghost number 
preserving condition on the string world sheets, naturally extended to  the 
polynomials of $\psi^i$ and $\phi^j$. 
Then $f(\V^{g,n}_{b,m})$ in eq.(\ref{eq5.7}), 
which is the image of $\ket{\Psi}^{\otimes n}\otimes 
\prod_{k=1}^b\ket{\Phi}^{\otimes m_k}$ 
by the map (\ref{mapOmega}), belongs to a subspace of $C(\psi,\phi)$ 
whose elements are expressed in general in the form 
\begin{equation*}
 a(\phi,\psi)=
 \sum_{k,l}\ov{k!l!|J_1|\cdots|J_l|} 
 a_{i_1\cdots i_k;J_1,\cdots,J_l}
 (\phi^{J_l}\cdots\phi^{J_1})(\psi^{i_k}\cdots\psi^{i_1})\ .
\end{equation*}
{}For the interaction terms, $|J_i|=m_i$ 
in the notation in eq.(\ref{eq5.7}). 
Here $J=(j_1,\cdots, j_{|J|})$ is the multi-index, 
$\phi^{J}=\phi^{j_{|J|}}\cdots\phi^{j_1}$. 
The coefficient $a_{i_1\cdots i_k;J_1,\cdots,J_l}\in\C$ is then 
graded symmetric with respect to the cyclic permutations of each multi-index
$J=(j_1,\cdots,j_{|J|})$, 
all the permutations of $i_1\cdots i_k$, and those of $J_1,\cdots,J_l$. 
We denote the corresponding subspace by 
$C(\phi,\psi)_{qoc}:=C_{sym}(\psi)\otimes C_{sym}(C(\phi)_{cyc})\subset
C(\phi,\psi)$, where `$qoc$' indicates  `quantum open-closed'. 
Note that, by construction, 
the degree of $f(\V^{g,n}_{b,m})$ is zero. 
The closed string kinetic term and the open string kinetic term 
are expressed as follows, 
\begin{equation}
 \half \langle \Psi, Q_c \Psi\rangle_c\ ,\qquad 
 \half \langle \Phi, Q_o \Phi\rangle_o
 \label{oc-kinetic}
\end{equation}
which also belong to $C(\phi,\psi)_{qoc}$ and have degree zero. 
In our notation, $Q_c=l_1:\cH_c\raw\cH_c$ and 
$Q_o=n_{0,1}:\cH_o\raw\cH_o$. 
Physically, $Q_c$ (resp. $Q_o$) is called the {\em BRST operator} for 
closed (resp. open) strings, where BRST is taken in the sense of 
the conformal field theory on the string world sheet. 
Their cohomologies then define the physical state spaces of strings. 
Also, the brackets are just the constant symplectic structures 
in Definition \ref{defn:sym}: 
\begin{equation*}
 \langle\ \ ,\ \ \rangle_c=\omega_c: \cH_c\otimes\cH_c\raw\C\ ,
 \qquad 
  \langle\ \ ,\ \ \rangle_o=\omega_o: \cH_o\otimes\cH_o\raw\C\ .
\end{equation*}
Since these constant symplectic structures come 
from the BV-formalism \cite{BV1,BV2} in which string field theories 
are described, 
the degrees of $\omega_c$ and $\omega_o$ are set 
to be minus one. 
In such a superfield description of the BV-formalism, 
they are called {\em odd symplectic structures} \cite{Sch,AKSZ} 
since `degree minus one' implies {\em odd} in $\Z_2$ grading.
The corresponding odd Poisson brackets 
\begin{equation*}
 (\ ,\ )=(\ ,\ )_c+(\ ,\ )_o\ ,\qquad 
 (\ ,\ )_c=\flpartial{\psi^i}\omega_c^{ij}\frpartial{\psi^j}\ ,
\quad 
 (\ ,\ )_o=\flpartial{\phi^i}\omega_o^{ij}\frpartial{\phi^j}\ 
\end{equation*}
are what are called the {\em BV-brackets}. 
Since they have degree one, 
$(\psi^i,\psi^j)_c=\omega_c^{ij}\ne 0$ only when 
the sum of the degree of $\psi^i$ and the degree of $\psi^j$ is 
equal to minus one. 
In the BV-formalism \cite{BV1,BV2}, 
two fields $\psi^i$ and $\psi^j$ having nonzero $\omega_c^{ij}$ 
make a pair of a {\em field} and an {\em antifield}. 
Of course these facts hold true similarly for open string fields 
(see \cite{Ka,thesis}). 
Both Poisson brackets are naturally combined  and 
extended to $(\ ,\ )$ on 
$C(\phi,\psi)_{qoc}$. 
Also, define second order operators as 
\begin{equation}
 \Delta=\Delta_c+\Delta_o\ ,\qquad 
 \Delta_c=\half(-1)^{\eb_{c,i}}
 \omega_c^{ij}\frpartial{\psi^i}\frpartial{\psi^j}\ ,\quad 
  \Delta_o=\half(-1)^{\eb_{o,i}}
 \omega_o^{ij}\frpartial{\phi^i}\frpartial{\phi^j}\ .  
\end{equation}
Since the BV-brackets have degree one, we have $\deg(\Delta)=1$, while 
$(C(\psi),\Delta_c,\bullet, (\ ,\ )_c)$ and 
$(C(\phi)_{cyc},\Delta_o,\bullet, (\ ,\ )_o)$ 
form BV-algebras (see \cite{BV1,BV2,get1}). 

Further 
$(C(\psi,\phi)_{qoc},\Delta,\bullet, (\ ,\ ))$ 
forms a BV-algebra, where 
$\bullet: C(\psi,\phi)_{qoc}\otimes C(\psi,\phi)_{qoc}\raw C(\psi,\phi)_{qoc}$ 
is the associative product, symmetric in the  
$\psi$'s. 
We shall soon reduce them to the tree open-closed structures, so 
do not stress to explain the detail of the structure on the whole 
$C(\psi,\phi)_{qoc}$ in this paper. 

The action of quantum open-closed string field theory 
is then given by summing up the kinetic terms (\ref{oc-kinetic}) 
and all the interaction terms (vertices) in eq.(\ref{eq5.7}): 
\begin{equation}\label{S_qoc}
 \begin{split}
 & S_{qoc}(\phi,\psi)=\sum_{g,b,n}
 \ov{n!b!|J_1|\cdots|J_b|} \V^g_{i_1\cdots i_n;J_1,\cdots,J_b}
(\phi^{J_b}\cdots\phi^{J_1})(\psi^{i_n}\cdots\psi^{i_1})\ ,\\
 & \V^g_{i_1\cdots i_n;J_1,\cdots,J_b}
 :=\int_{\V^{g,n}_{b,m}}\bra{\Omega}
(\eb_{c,i_1},\cdots,\eb_{c,i_n};\eb_{o,J_1},\cdots,\eb_{o,J_b})\ ,
 \end{split}
\end{equation}
where $\eb_{o,J}=\eb_{o,j_1}\cdots \eb_{o,j_{|J|}}$ 
for $J=(j_1,\cdots j_{|J|})$, and 
the summation $\sum_{g,b,n}$ is taken for all 
$g\ge 0$, $b\ge 0$ and $n\ge 0$ except for the cases 
$(g,b)=(0,0), n\le 1$, $(g,n,b)=(0,0,1), |J_1|\le 1$ and 
$(g,n,b)=(1,0,0)$. 
In particular, the terms with $(g,n,b)=(0,2,0)$ and 
$(g,n,b)=(0,0,1)$ with $|J_1|=2$ are the kinetic terms of 
closed strings and open strings, respectively. 

A quantum open-closed string field theory $S_{qoc}(\phi,\psi)$ 
is defined so that it satisfies the {\it quantum master equation}
\begin{equation}
 \half(S_{qoc},S_{qoc})+ \Delta S_{qoc}=0\ .
\end{equation}
Note that $\Delta S_{qoc}$ is the term peculiar to the {\em quantum} 
string field theory. 
$\Delta_c$ increases $g$ by one and 
$\Delta_o$ increases $b$ by one for $b>0$. 
The quantum master equation splits into separate equations
for each genus $g$ and number of  boundaries $b$. 
When we concentrate on the equations for $g=0$ and $b=0$ or $1$, 
we get 
\begin{align}
 (g,b)=(0,0)\ , &\quad 0=(S_S,S_S)_c\ , \label{mastereq0}\\
 (g,b)=(0,1)\ , &\quad 0=(\ti{S}_D,S_S)_c+\half(\ti{S}_D,\ti{S}_D)_o \ , 
 \label{mastereq1}
\end{align}
where $S_S$ and $\ti{S}_D$ consist of the corresponding terms in 
$S_{qoc}$ in eq.(\ref{S_qoc}); explicitly, 
$S_S$ is of the same form as $S_S$ in eq.(\ref{action2}) and 
$\ti{S}_D$ consists of $S_D$ in eq.(\ref{action2}) with additional 
terms corresponding to $(k,l)=(k,0)$ below: 
\begin{equation}
 \begin{split}
& S_S(\psi)
=\sum_{l\ge 2}\ov{l!}\V_{i_1\cdots i_l}\psi^{i_l}\cdots\psi^{i_1}
 \ \in C(\psi)\ , \\
& \ti{S}_D(\phi,\psi)
 =\sum_{k\ge 0,l\ge 0, 2k+l\ge 2}
 \ov{k!\ l}\V_{i_1\cdots i_k;j_1\cdots j_l}(\phi^{j_l}\cdots\phi^{j_1})
 (\psi^{i_k}\cdots\psi^{i_1})\ \in C(\phi,\psi)_{cyc}\ . 
 \end{split}
 \label{action-ocsft}
\end{equation}
Here we dropped the index $g$ 
used in eq.(\ref{S_qoc}) 
since $g=0$. Namely, we denote
\begin{equation*}
 \V^{g=0}_{i_1\cdots i_l;\emptyset}=:\V_{i_1\cdots i_l}\ ,\qquad 
 \V^{g=0}_{i_1\cdots i_l;J=(j_1,\cdots,j_l)}
 =:\V_{i_1\cdots i_l;j_1\cdots j_l}\ . 
\end{equation*}
The action $S_S$ corresponds to punctured spheres 
(since the corresponding Riemann surfaces have no boundary
($\emptyset$)), 
whereas $\ti{S}_D$ corresponds to disks with punctures 
both in the disks and on the boundary of the disks. 
Equation (\ref{mastereq1}) is often called 
a {\em Maurer-Cartan equation}. 

A classical (tree) open-closed string field theory \cite{Z2} 
is then given by the action 
$S_{toc}(\phi,\psi)=S_S(\psi)+ \ti{S}_D(\phi,\psi)$ 
satisfying eq.(\ref{mastereq0}) and (\ref{mastereq1}), 
the Batalin-Vilkovisky \cite{BV1,BV2}
classical open-closed master equations. 
The identity (\ref{mastereq0}) implies that 
$S_S$ is just the action of 
the classical closed string field theory \cite{Z1}. 
Namely, $S_S$ has a cyclic $L_\infty$-structure. 
For the operadic construction of the classical closed string field
theory, see \cite{KSV}. 
The relevant operad is the $L_\infty$-operad of non-planar trees, 
where the composition of the trees corresponds to the 
{\em twist-sewing} of two $S^1$'s parametrizing 
two closed strings/boundaries 
in a Riemann surface picture \cite{Z1,KSV}.

Just in the same way as for eq.(\ref{fvf}), one can define 
the following formal vector fields acting on 
$C(\phi,\psi)_{cyc}$, 
\begin{equation}
 \delta:=\delta_S+\delta_D\ ,\qquad 
 \delta_S=(\ ,S_S)_c\ ,\quad\delta_D=(\ ,\ti{S}_D)_o \ .
 \label{fvf2}
\end{equation}
The condition $(\delta)^2=0$ that $(\cH,\l,\n)$ is a
cyclic OCHA is 
equivalent to the derivatives of the master equations 
(\ref{mastereq0}),(\ref{mastereq1}): 
\begin{align*}
 & 0=\(\quad ,(S_S,S_S)_c\)_c\ ,\\
 & 0=\(\quad\ ,(\ti{S}_D,S_S)_c+\half(\ti{S}_D,\ti{S}_D)_o\)_o\ .
\end{align*} 
Here note that, as has been explained in eq.(\ref{action-ocsft}), 
$\ti{S}_D$ consists of $S_D$ with the following additional terms 
\begin{equation}
\ov{l!}\V_{j_1\cdots j_l;J=\emptyset}(\psi^{j_l}\cdots\psi^{j_1})\ 
 \label{const-term}
\end{equation}
each of which corresponds to a disk with punctures only in the bulk 
(the interior of the disk) 
and no punctures $J=\emptyset$ on the boundary. 
However, one can see that these terms drop out in eq.(\ref{fvf2}): 
$(\ ,\ti{S}_D)_o=(\ ,S_D)_o$, 
since no open string field $\phi^i$ is included in eq.(\ref{const-term}). 
This is why we do not include the corresponding terms 
in the definition of (cyclic) OCHAs. 
Thus, a classical open-closed string field theory is 
a cyclic OCHA with the additional terms (\ref{const-term}). 

Of course, there exist situations in which these terms 
(\ref{const-term}) are also 
important physically. For example, 
the terms (\ref{const-term}) contribute to 
a constant term for open string field theory 
in discussing its deformation 
as in section \ref{sec:def}, and the constant term is 
relevant to D-brane mass, 
since the value of the action is believed to correspond 
to D-brane mass in open string field theory. 
But, it is enough to consider a cyclic 
OCHA structure in a classical open-closed string field 
theory at present if we examine its homotopy algebraic structures 
in the sense of the next subsection.

\section{Maurer-Cartan equation, minimal model and 
tree open-closed string amplitudes}
\label{sec:MMth}

Homotopy algebras should have some homotopical properties 
\cite{BoVo,markl:haha}. 
One of the key theorems in homotopy algebra is the minimal model
theorem. The minimal model theorem for $A_\infty$-algebras was proved by 
Kadeishvili \cite{kadei1}. 
For the construction of minimal models of $A_\infty$-structures,
homological perturbation theory was developed by 
\cite{gugen,hueb-kadei,gugen-lambe,GS,GLS:chen,GLS}, 
for instance, 
and the form of a minimal model is then given explicitly 
in \cite{mer,KoSo}. Also, the existence of a stronger theorem, 
called the decomposition theorem in \cite{thesis,KaTe}, is mentioned 
in \cite{Ko1} and proved by employing 
a kind of homological perturbation theory 
in \cite{thesis,KaTe} (see also \cite{Le-Ha}). 
It is clear that the same arguments hold true for $L_\infty$-algebras, 
and in our previous paper \cite{OCHA} 
we stated that they hold also for OCHAs. 

In this section, we present the explicit form of a minimal model 
for an OCHA, which in the cyclic case can be thought of as the 
perturbative expansion of a classical open-closed string field theory. 
\begin{defn}[Minimal open-closed homotopy algebra]
An OCHA $(\cH=\cH_c\oplus\cH_o,\l,\n)$ 
is called minimal if 
$l_1=0$ on $\cH_c$ and $n_{0,1}=0$ on $\cH_o$. 
 \label{defn:minimalalg}
\end{defn}
\begin{thm}[Minimal model theorem for open-closed homotopy algebras 
\cite{OCHA}]
For any OCHA, there exists a minimal OCHA 
and an OCHA quasi-isomorphism from the minimal OCHA 
to the original OCHA. 
 \label{thm:minimal}
\end{thm}
The minimal model theorem holds also for COCHAs. 
Namely, for any COCHA, there exists a minimal COCHA 
and a COCHA quasi-isomorphism from the minimal COCHA 
to the original COCHA. 
This fact also follows from the explicit minimal model 
we shall construct here. 

First of all, we fix a Hodge decomposition of the complex 
$(\cH,d=l_1+n_{0,1})$.
Namely, for $d_c = l_1$ and $d_o = n_{0,1}$, 
we give Hodge decompositions of the complexes 
$(\cH_c,d_c)$ and $(\cH_o,d_o)$ separately, 
by fixing degree minus one (homotopy) operators 
$h_c:\cH_c\raw\cH_c$ and $h_o:\cH_o\raw\cH_o$:  
\begin{equation}
 d_c h_c+h_c d_c + \iota_c\circ\pi_c=\1_c\ \ ,\qquad 
 d_o h_o+h_o d_o + \iota_o\circ\pi_o=\1_o\ \ .
 \label{HKdecomp}
\end{equation}
Here, $\pi$ and $\iota$ indicate the projection to and the inclusion 
into the corresponding cohomologies. 
Thus, these data give a contraction (deformation retract) of 
$\cH=\cH_c\oplus\cH_o$ as a graded vector space (see \cite{OCHA}); 
\begin{equation}
 \Nsddata {H(\cH)} {\hspace*{0.35cm}\iota}{\hspace*{0.35cm}\pi}{\cH}h\ ,
 \label{SDR}
\end{equation}
where $\iota:=\iota_c+\iota_o$, 
$\pi:=\pi_c+\pi_o$ and $h=h_c+h_o$. 

We would like to follow the arguments in \cite{Ka,thesis}, 
where a minimal model is obtained by a process of 
solving the Maurer-Cartan equation for an $A_\infty$-algebra. 
For OCHAs, the Maurer-Cartan equations are defined as follows 
\cite{OCHA}. 
\begin{defn}[Maurer-Cartan equation] 
For an OCHA $(\cH,\l,\n)$ 
and degree zero elements $\cb\in\cH_c$ and $\ob\in\cH_o$, 
we call the following pair of equations 
\begin{equation}
0=\sum_k \ov{k!}l_k(\cb,\cdots,\cb)\ ,\qquad 
0=\sum_{k,l}\ov{k!} 
 n_{k,l}(\cb,\cdots,\cb;\ob,\cdots,\ob)\ 
 \label{mceq}
\end{equation}
the {\em Maurer-Cartan equations} for the OCHA 
$(\cH,\l,\n)$. 
 \label{defn:mceq}
\end{defn}

\begin{rem}
Recall that, for the cyclic $A_\infty$ or $L_\infty$ case, 
the Maurer-Cartan equations are just the equations of motions for 
the action (of the corresponding string field theory)
\cite{Ka,thesis}. 
In field theory, the equations of motions are defined by the derivatives 
of the action with respect to the fields. 
For instance, for classical closed string field theory with 
the action $S_S$, the equations of motions are $0=\frpartial{\psi^i}S_S$ 
for each $i$. Here, since the BV-bracket 
$(\ ,\ )_c:=\flpartial{\psi^i}\omega^{ij}_c\frpartial{\psi^j}$ 
is nondegenerate, the equations of motions are equivalent to 
$0=(\ ,S_S)\ (=\delta_S)$. Usually, we set degree nonzero fields to be
zero and concentrate on the solutions for the fields of degree zero. 
Then, further identifying $\frpartial{\psi^i}$ with $\eb_c^i$ in 
$0=\delta_S,$ 
one obtains the Maurer-Cartan equations for the $L_\infty$-algebra, 
which is the first equation in eq.(\ref{defn:mceq}). 
However, note that, for a COCHA $(\cH,\omega,\l,\n)$, 
the zeroes of the corresponding odd formal vector field 
$\delta=\delta_S+\delta_D$ 
are not the same as solutions to the equations of motions 
$0=(\ ,S)=(\ ,S)_c+(\ ,S)_o$,
or separately 
\begin{equation*}
 0=\flpartial{\psi^i}\omega_c^{ij}\frpartial{\psi^j}(S_S+S_D)=0\
 ,\qquad 
 0=\flpartial{\phi^i}\omega_o^{ij}\frpartial{\phi^j}S_D=0\ , 
\end{equation*}  
for the COCHA. 
Namely, the first equation above includes the term $(\ ,S_D)_c$, 
the corresponding term of which is absent in the 
Maurer-Cartan equations (\ref{defn:mceq}) for the COCHA. 

One can see that, 
if one solves the equations of motions, 
instead of solving the Maurer-Cartan equations, 
the resulting structure includes terms corresponding to $b > 1$. 
 \label{rem:mc-eom}
\end{rem}
If we apply the arguments in \cite{Ka,thesis} to an OCHA, 
the Maurer-Cartan equations for an OCHA should be considered formally 
for the pair of `string fields' 
$(\Psi,\Phi)\in (\cH_c\otimes\cH_c^*,\cH_o\otimes\cH_o^*)$ 
instead of their degree zero parts 
$(\cb,\ob)\in (\cH_c^0,\cH_o^0)$.

Then, for instance for the $L_\infty$ part, solving the Maurer-Cartan 
equation recursively one gets first an $L_\infty$-quasi-isomorphism 
$\{f_k\}_{k\ge 1}: (H(\cH_c))^{\otimes k}\raw\cH_c$. 
This somewhat physical procedure is closely related to the 
homological perturbation theory developed earlier, 
and in particular, 
$\f:=\oplus_k f_k\in \oplus_k \Hom((H(\cH_c))^{\otimes k},\cH_c)$ 
is just what is called a {\em twisting cochain} $\tau$ 
(see \cite{jh-jds} for the DGLA case). 
Then, substituting $\f$ instead of $\Psi$ 
into the initial Maurer-Cartan equation, 
one obtains 
an equation on $H(\cH_c)$, which is in fact 
the Maurer-Cartan equation for the corresponding 
minimal $L_\infty$-algebra, so one can read 
the minimal $L_\infty$-structure from the Maurer-Cartan equation. 
For the case of an OCHA, 
its minimal model is obtained by first considering 
the Maurer-Cartan equation for the $L_\infty$-algebra as above 
and, after that, considering the Maurer-Cartan equation for $\n$. 

For an $L_\infty$-algebra $(\cH_c,\l)$, 
a minimal $L_\infty$-algebra 
and an $L_\infty$-quasi-isomorphism 
$\{f_l\}_{l\ge 1}:(H(\cH_c),\l')\raw(\cH_c,\l)$ 
are constructed as follows. 

We set $f_1=\iota_c:H(\cH_c)\raw\cH_c$, and 
assume that we have $\{f_l:(H(\cH_c))^{\otimes l}\raw\cH_c\}_{l\ge 1}$ 
for $l\le n-1$. 
Then, for $c'_1,\cdots, c'_n\in H(\cH_c)$, 
$f_n$ is defined by 
\begin{equation*}
f_n(c'_1,\cdots,c'_n)
 =-h_c\sum_{k_1+\cdots +k_j =n}\,
\sum_{\sigma\in\S_n}
\frac{(-1)^{\epsilon(\sigma)}}{k_1!k_2!\cdots k_j!\cdot j!} 
l_j(f_{k_1}\otimes f_{k_2}\otimes \cdots \otimes f_{k_j})
(c'_{\sigma(1)},\cdots,c'_{\sigma(n)})\ .
\end{equation*}
The minimal $L_\infty$-structure is then determined by 
\begin{equation*}
 l'_n(c'_1,\cdots,c'_n)
 =\pi_c\sum_{k_1 +\cdots +k_j =n}\, \sum_{\sigma\in\S_n}
\frac{(-1)^{\epsilon(\sigma)}}{k_1!k_2!\cdots k_j!\cdot j!} 
l_j(f_{k_1}\otimes f_{k_2}\otimes\cdots\otimes f_{k_j})
 (c'_{\sigma(1)},\cdots,c'_{\sigma(n)})\ ,
\end{equation*}
in particular, for $l=2$ one gets 
$l'_2=H(l_2):=\pi_c\circ l_2\circ (\iota_c)^{\otimes 2}$.

\def \ccdots{\scalebox{0.8}[1]{\text{$\cdots$}}}

Once the $L_\infty$-quasi-isomorphism $\{f_l\}_{l\ge 1}$ is 
given, 
we have an OCHA $(H(\cH_c)\oplus\cH_o,\l',\n'')$ for some $\n''$. 
Next we should construct 
$\{
 f_{k,l}:(H(\cH_c))^{\otimes k}\otimes(H(\cH_o))^{\otimes l}\raw H(\cH_o)
\}_{k+l\ge 1}$ and 
$\n'=\{
n'_{k,l}:(H(\cH_c))^{\otimes k}\otimes(H(\cH_o))^{\otimes l}\raw H(\cH_o)
\}_{2k+l\ge 2}$; 
these are obtained 
in a similar way as follows.
$f_{0,1}$ is given as inclusion 
$f_{0,1}=\iota_o:H(\cH_o)\raw\cH_o$. 
$f_{r,s}$ and $f_{p,q}$ are ordered as 
$f_{r,s}\prec f_{p,q}$ if 
$r+s< p+q$ or $r<p$ for $r+s=p+q$. 
Then, a similar recursive procedure as above 
can be carried out also here. 
For $c'_1,\cdots,c'_n\in H(\cH_c)$ and $o'_1,\cdots,o'_m\in H(\cH_o)$, 
$f_{n,m}$ is determined by 
\begin{equation*}
 \begin{split}
& f_{n,m}(c'_1,\cdots,c'_n;o'_1,\cdots,o'_m) = \\ &\ -h_o
\sum 
\frac{(-1)^{\epsilon(\sigma)}}
{i!(r_1!\ccdots r_i!)(p_1!\ccdots p_j!)} 
n_{i,j}(f_{r_1}
\otimes\ccdots\otimes f_{r_i}
\otimes f_{p_1, q_1}
\otimes\ccdots\otimes f_{p_j,q_j})
(c'_{\sigma(1)},\cdot\cdot,c'_{\sigma(n)};o'_1,\cdot\cdot,o'_m)\ , 
 \end{split}
\end{equation*}
where the summation $\sum$ runs over 
all $r_1,\cdots,r_i, p_1,\cdots,p_j,q_1,\cdots,q_j$ such that 
$(r_1+\ccdots +r_i)+(p_1+\ccdots +p_j)=n$, 
$(q_1+\ccdots +q_j)=m$, and also all $\sigma\in\S_n$. 

Then $\n'=\{n'_{k,l}\}_{2k+l\ge 2}$ 
is obtained by replacing $-h_o$ above with $\pi_o$, 
\begin{equation*}
 \begin{split}
& n'_{n,m}(c'_1,\cdots,c'_n;o'_1,\cdots,o'_m) =\\ &\ \pi_o
\sum 
\frac{(-1)^{\epsilon(\sigma)}}
{i!(r_1!\ccdots r_i!)(p_1!\ccdots p_j!)} 
n_{i,j}(f_{r_1}
\otimes\ccdots\otimes f_{r_i}
\otimes f_{p_1, q_1}
\otimes\ccdots\otimes f_{p_j,q_j})
(c'_{\sigma(1)},\cdot\cdot,c'_{\sigma(n)};o'_1,\cdot\cdot,o'_m)\ , 
 \end{split}
\end{equation*}
where the summation $\sum$ stands for the same one as above. 
In particular, for $2k+l=2$ one gets 
$n'_{0,2}=H(n_{0,2}):=\pi_o\circ n_{0,2}\circ(\iota_o)^{\otimes 2}$ and 
$n'_{1,0}=H(n_{1,0}):=\pi_o\circ n_{1,0}\circ\iota_c$. 
In the equation above, we used the convention presented in 
Definition \ref{defn:ocmorp}. 

{}{}{}
For a COCHA $(\cH,\omega,\l,\n)$, 
we do this construction 
by starting with an {\em orthogonal} Hodge decomposition 
with respect to the symplectic form $\omega$. Namely, 
we give a decomposition 
$$\Nsddata {H(\cH)}
{\hspace*{0.35cm}\iota}{\hspace*{0.35cm}\pi}{\cH}h$$ 
of $\cH$ in eq.(\ref{SDR}) with a homotopy $h:\cH\to\cH$ 
satisfying $\omega(\1\otimes h)=\omega(h\otimes \1)$, 
where $\1:=\1_c\oplus\1_o$. 
The existence of such a homotopy $h$ follows from the 
nondegeneracy of $\omega$ and the 
cyclicity for the terms $\omega_c(l_1\otimes\1_c)$ and 
$\omega_o(n_{0,1}\otimes\1_o)$, 
and then 
$\omega(\1\otimes(\iota\circ\pi))=\omega((\iota\circ\pi)\otimes\1)$ 
also holds. 
Then, for the COCHA $(\cH,\omega,\l,\n)$, 
forgetting the cyclic structure $\omega$ having already
used it to fix the contraction (\ref{SDR}), 
one can obtain a minimal model $(H(\cH),\l',\n')$ as an OCHA 
by the construction we have seen above. 
The resulting minimal 
OCHA $(H(\cH),\l',\n')$ is in fact cyclic with respect to the 
induced inner product $\omega':=\omega(\iota\otimes\iota)$ and 
the OCHA quasi-isomorphism a COCHA quasi-isomorphism. 

To summarize: 
\begin{thm}
$(H(\cH),\l',\n')$ forms a minimal OCHA and 
$\f:=\{\{f_l\}_{l\ge 1}, \{f_{k,l}\}_{k+l\ge 1}\}$ is 
an OCHA quasi-isomorphism $\f: (H(\cH),\l',\n')\raw (\cH,\l,\n)$. 
 \label{thm:min-construct}
\end{thm}
\begin{thm}
For a COCHA $(\cH,\omega,\l,\n)$ and 
an orthogonal Hodge decomposition with respect to $\omega$, 
$(H(\cH),\omega',\l',\n')$ forms a minimal COCHA and 
$\f:=\{\{f_l\}_{l\ge 1}, \{f_{k,l}\}_{k+l\ge 1}\}$ is 
a COCHA quasi-isomorphism $\f: (H(\cH),\omega',\l',\n')\to
(\cH,\omega,\l,\n)$. 
 \label{thm:cyc-min-construct}
\end{thm}
Since the explicit forms are given, 
one can check the cyclicity directly 
in a similar way to that in the $A_\infty$ case (see \cite{thesis}). 
\begin{rem}[Rooted planar tree graphs]
One can also present an alternate description of this minimal model 
in terms of rooted planar trees in a similar way as for 
$A_\infty$-algebras (see \cite{KoSo,Ka,thesis,markl:transfer}, etc.). 
This is related to Feynman graphs in field theory. 
For an $A_\infty$-algebra $(\cH_o,\{n_{0,k}\})$, 
it is convenient to associate $n_{0,k}$ to the $k$-corolla 
of planar rooted trees. 
An $L_\infty$-algebra also has such a description, where 
the $L_\infty$-structure $l_k$ is associated with the $k$-corolla of 
nonplanar rooted trees (Figure \ref{fig:corolla1}). 
\begin{figure}
 \hspace*{1.0cm}
 \includegraphics{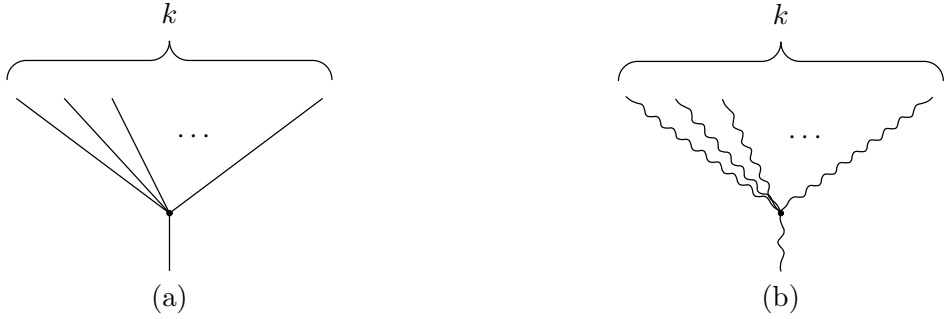}
 \caption{(a). The $k$-corolla, the planar tree corresponding to the 
$A_\infty$-structure $m_k=n_{0,k}$. 
(b). The $k$-corolla, the nonplanar tree corresponding to the 
$L_\infty$-structure $l_k$. }
 \label{fig:corolla1}
\end{figure}
In our OCHA, we need to introduce also 
$n_{k,l}$, to which we associate a `mixed-corolla ' as in 
Figure \ref{fig:corolla2}. 
\begin{figure}
 \hspace*{1.0cm}
 \includegraphics{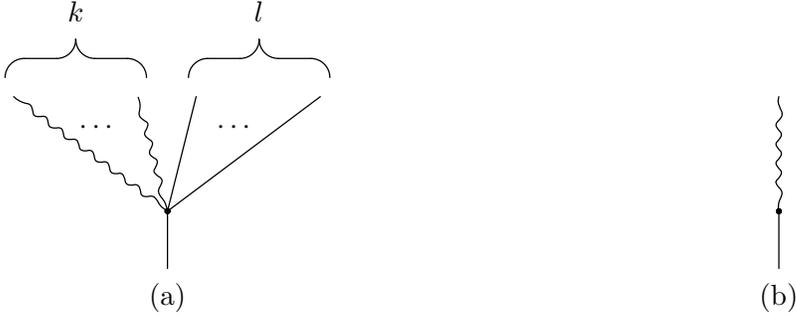}
 \caption{(a). The $(k,l)$-corolla corresponding to $n_{k,l}$. 
(b). For the open-closed case, the $(k,l)$-corollas include the 
$(1,0)$-corolla corresponding to $n_{1,0}$. }
 \label{fig:corolla2}
\end{figure}
As stated previously, from a string theory viewpoint, 
$l_k$ corresponds to 
the sphere with $(k+1)$-(closed string) punctures and 
$n_{k,l}$ corresponds to 
the disk with $(k+1)$-(open string) punctures on the boundary and 
$l$-(closed string) punctures in the bulk (interior) of the disk. 
In fact, one may think of the tubular neighborhood of these tree graphs 
as the corresponding world sheet, 
where we take strips and cylinders for the neighborhood of 
the straight lines and meandering lines, respectively. 
The minimal OCHA structure 
$l'_k$ and $n'_{k,l}$ are then obtained by grafting 
corollas in all possible ways such that straight lines are 
grafted to straight and wiggly to wiggly (Figure \ref{fig:ww-ss}), 
where we assign to corollas the corresponding 
multi-linear maps $l_k,n_{p,q}$, and to internal edges 
$h_c$, $h_o$, and so on. 
\begin{figure}[h]
\hspace*{3.5cm}
\begin{minipage}[c]{50mm}
{\includegraphics{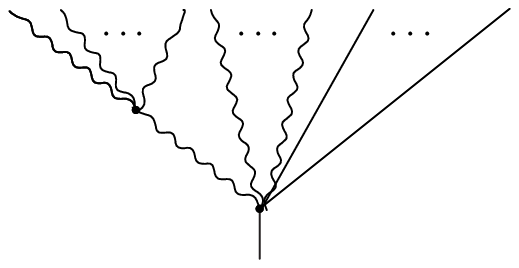}}\end{minipage}
 \qquad 
\begin{minipage}[c]{50mm}
{\includegraphics{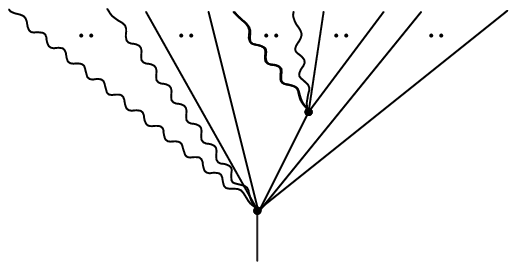}}\end{minipage}

\vspace*{2.2cm}

\hspace*{5.0cm}(a)\hspace*{5.7cm}(b)
 \caption{Grafting of corollas.\ 
(a): Wiggly to wiggly. (b): Straight to straight. }
 \label{fig:ww-ss}
\end{figure}
Physically, $h_c$ and $h_o$ are the propagators for closed string 
and open string, respectively. 
 \label{rem:minimal2}
\end{rem}
\begin{rem}[String amplitude]
For a classical open-closed string field theory 
$S=S_S+S_D$, the string amplitudes are obtained as follows. 
Let $(\cH,\omega,\l,\n)$ be the corresponding COCHA, and 
suppose that its minimal COCHA $(H(\cH),\omega',\l',\n')$ 
is constructed as above. 

By definition, string field theory is constructed so that 
its perturbative expansion reproduces the corresponding 
world sheet string amplitudes. 
Thus, 
\begin{equation}
 \V'_{k+1}:=\omega'_c(l'_k\otimes\1_c)\ ,\qquad 
 \V'_{k,l+1}=\omega'_o(n'_{k,l}\otimes\1_o)\ 
\end{equation} 
just define the on-shell 
$(k+1)$-closed strings sphere amplitudes and 
$k$-closed $(l+1)$-open string disk amplitudes, respectively. 
Moreover, the $n$-closed string disk amplitude, 
which we denote $\V'_{n,0}$, is given by composing 
the $L_\infty$-quasi-isomorphism with $\V_{k,0}$ as follows:
\begin{equation*}
 \V'_{n,0}=\sum_{i=1}^n\ \sum_{k_1+\cdots+k_i=n}
 \ov{k_1!\cdots k_i!}
 \V_{i,0}(f_{k_1}\otimes\cdots\otimes f_{k_i})
\end{equation*}
for $\f=\{f_k\}_{k\ge 1},$ the $L_\infty$-quasi-isomorphism. 
 \label{rem:stg-amp}
\end{rem}
Note also that we can prove the decomposition theorem 
for a (cyclic) OCHA \cite{OCHA}, which implies that 
all classical open-closed string field theories constructed on 
a fixed {\em conformal background} (the data of the 
conformal field theory on the Riemann surfaces) are isomorphic. 
{}From the field theory viewpoint, 
an OCHA-morphism is a field transformation and in particular 
an OCHA-isomorphism is a field redefinition. 
Since field theory actions related by a field redefinition are 
physically equivalent, 
one can say the decomposition theorem can prove the equivalence of 
all classical open-closed string field theories constructed on 
a fixed conformal background. 
(See \cite{thesis} for the $A_\infty$ case.)

 \section{Deformation of $A_\infty$-structures and the formality theorem}
\label{sec:def}

For an $L_\infty$-algebra $(\cH_c,\l)$ and an $A_\infty$-algebra 
$(\cH_o,\m)$, if there exists 
an OCHA $(\cH=\cH_c\oplus\cH_o,\l,\n)$ whose 
sub-algebra $(\cH_o,\{n_{0,k}\}_{k\ge 1})$ coincides with 
$(\cH_o,\{m_k\}_{k\ge 1})$, 
one obtains deformations of the $A_\infty$-algebra
$(\cH_o,\{m_k\}_{k\ge 1})$ parameterized 
by the $L_\infty$-algebra $(\cH_c,\l)$. 
On the other hand, the whole deformation space of the $A_\infty$-algebra 
is also described by a moduli space of
the $L_\infty$-algebra $(\Coder(T^c\cH_o),\m,[\ ,\ ]))$, 
which we denote by $(\cH_c',\l').$ The maps
$\{n_{k,l}\}$ for $k\ge 1$ define an $L_\infty$-morphism from 
$(\cH_c,\l)$ to $(\cH_c',\l')$. 
The defining equation for an OCHA (\ref{occd})
just converts to the defining equation for an $L_\infty$-morphism
(\ref{Linftymorp}) \cite{OCHA}. 
In subsection \ref{ssec:def} we re-explain this 
in a more explicit way. 
Such structure appears in various aspects of mathematical physics; 
the relation to deformation quantization by Kontsevich \cite{Ko1} is 
explained in subsection \ref{ssec:K}, and the application to 
the open-closed topological B-model \cite{Hof2} from the viewpoint of the
homological mirror symmetry set-up \cite{mirror} is discussed in 
subsection \ref{ssec:Bmodel}. 
Note also that, in string field theory, 
this picture is related to the arguments in section 8 of \cite{Z2}.

 \subsection{Deformations of $A_\infty$-structures 
from open-closed homotopy algebras}
\label{ssec:def}

\begin{defn}[(Graded) Gerstenhaber bracket \cite{gerst}]
For $\cH_o$ a $\Z$-graded vector space, 
let $\Hom^r_k:=\Hom^r(\cH_o^{\otimes k},\cH_o)$ be the space 
of degree $r$ $k$-linear maps, and 
\begin{equation*}
 \Hom:=\oplus_r \Hom^r\ ,\qquad 
 \Hom^r:=\oplus_{k\ge 0}\Hom^r_k\ .
\end{equation*}
It is known that $\Hom$ is in one-to-one correspondence 
with the space of coderivations 
on $T^c(\cH_o)$, $\Coder(T^c(\cH_o))$ \cite{jds:intrinsic}. 
For two elements in $\Coder(T^c(\cH_o))$, the commutator 
of any two elements in fact belongs to $\Coder(T^c(\cH_o))$, and further 
satisfies the Jacobi identity. 
After a shift in degree, this induces a graded Lie bracket 
on $\Hom$, which is the {\em Gerstenhaber bracket}. 
For $m\in \Hom_k^r$, $m'\in \Hom_{k'}^{r'}$ and 
$o_1,\cdots,o_{k+k'-1}\in\cH_o$, 
the graded Lie bracket $[m,m']\in \Hom_{k+k'-1}^{r+r'}$ is defined by 
\begin{equation*}
 \begin{split}
 &[m,m']=m\circ m'-(-1)^{r\cdot r'}m'\circ m\ ,\\
 &m\circ m'(o_1,\cdots\! ,o_{k+k'-1})
 =\sum_{i=0}^{k-1}(-1)^{l'(o_1+\cdots +o_i)}
 m(o_1,\cdots\!
 ,o_i,m'(o_{i+1},\cdot\cdot,o_{i+k'}),o_{i+k'+1},\cdots\! ,o_{k+k'-1})\ .
 \end{split}
\end{equation*}
Thus, $(\Hom, [\ ,\ ])$ forms 
a {\em graded Lie algebra} \cite{gerst,jds:intrinsic}. 
 \label{defn:bracket}
\end{defn}
In the supermanifold description in section \ref{sec:dual}, 
this Gerstenhaber bracket corresponds to the graded Lie bracket 
of formal vector fields on the corresponding formal supermanifold.

Furthermore, let us denote by 
${\bar m}\in\Hom^1$ the degree one element 
corresponding to a degree one coderivation in 
$(\cH_c',\l')=\Coder(T(\cH_o))$, 
that is, an $A_\infty$-structure on $\cH_o$. 
Then, it is clear that $[{\bar m},{\bar m}]=0$ holds and 
$(\Hom, d, [\ \ ,\ \ ])$ forms a DGLA with $d=[{\bar m}, \ \ ]$.  

\begin{rem}[Suspension]
A DGLA is described as an $L_\infty$-algebra through the suspension $s$. 
For the DGLA $(\Hom,d, [\ ,\ ])$, 
the suspension is a degree shifting operator 
$$
s:\Hom\raw \Hom[1]=:\cH_c'\ .
$$ 
By this, a degree $r$ element $m\in \Hom^r$ is mapped to be a degree 
$(r-1)$ element $s(m)\in \Hom[1]^{r-1}={\cH_c'}^{r-1}$. 
This actually converts the degree preserving bracket 
$[\ ,\ ]:\Hom\otimes \Hom\raw \Hom$ into a 
degree one bilinear map $l'_2:\cH_c'\otimes\cH_c'\raw\cH_c'$ defined by 
\begin{equation*}
 l'_2(s(m),s(m'))
 := s \left((-1)^r[m,m']\right)\ ,\qquad 
 m\in \Hom^r\ ,\quad m'\in \Hom^{r'}\ .
\end{equation*}
One can see that the graded anticommutativity 
$[m',m]=-(-1)^{rr'}[m,m']$ is in fact replaced by the graded
commutativity in eq.(\ref{gcomm}), 
$l'_2(s(m'),s(m))=(-1)^{(r-1)(r'-1)}l'_2(s(m),s(m'))$, where 
$l_1'$ is given simply by $l_1'=s d s^{-1}=d$. 
Then $(\cH_c',\l'=\{l'_1,l'_2,l'_3=l'_4=\cdots=0\})$ forms an 
$L_\infty$-algebra. 
 \label{rem:s}
\end{rem}
Now, let us express the open-closed multi-linear maps as 
\begin{equation*}
 n_{k,l}(c_1,\cdots,c_k;o_1,\cdots,o_l)
 =:\left(n^*_l(c_1,\cdots,c_k)\right)(o_1,\cdots,o_l)\ ,
\end{equation*}
where
$n^*_l(c_1,\cdots,c_k)$ belongs to 
$\Hom_k^{\deg(c_1)+\cdots+\deg(c_k)+1}$.   
By this, the second term of the OCHA relation (\ref{occd}) can be rewritten as 
\begin{equation*}
-(-1)^{(c_{\sigma(1)}+\cdots +c_{\sigma(l)}+1)}
 \half[n^*_{m+1-k}(c_{\sigma(1)},\cdots,c_{\sigma(l)}),
n^*_{k}(c_{\sigma(l+1)},\cdots,c_{\sigma(n)})]\ (o_1,\cdots,o_m)\ ,
\end{equation*}
which, acting further with the suspension $s$ on the equation above, yields 
$$
-\half l'_2(s(n^*_{m+1-k})(c_{\sigma(1)},\cdots,c_{\sigma(l)}),
s(n^*_k)(c_{\sigma(l+1)},\cdots,c_{\sigma(n)}))(o_1,\cdots,o_m)\ .
$$
Thus, one obtains 
\begin{equation}
\begin{split}
&\sum_{k,l\ge 0}
\sum_{p=0}^{m-k}\sum_{\sigma\in\S_n}
\half\frac{(-1)^{\epsilon(\sigma)}}{l!(n-l)!} 
l'_2(s(n^*_{m+1-k})(c_{\sigma(1)},\cdots,c_{\sigma(l)}),
s(n^*_k)(c_{\sigma(l+1)},\cdots,c_{\sigma(n)}))\\
&\ \ =\sum_{\sigma\in\S_n}\sum_{l=1}^n
\frac{(-1)^{\epsilon(\sigma)}}{l!(n-l)!}
s(n^*_m)(l_l(c_{\sigma(1)},\cdots,c_{\sigma(l)}),
c_{\sigma(l+1)}\cdots,c_{\sigma(n)})\ .
\end{split}
 \label{occd3}
\end{equation}
These are  just the defining equations for an $L_\infty$-morphism 
(\ref{Linftymorp}). 
By treating the $l=0$ and $l=n$ cases  separately 
in the first line of the equation above, it becomes just the condition that 
the collection of multi-linear maps 
$s(n^*_k):(\cH_c)^{\otimes *}\raw \Hom_k[1]$ forms 
an $L_\infty$-morphism from $(\cH_c,\l)$ to $(\cH_c',\l')$. 
Here, note that $l_1'=[\mb,\ \ ]$ and $\mb=\sum_{l\ge 0}m_k$, 
the $A_\infty$-structure included in 
$(\cH,\l,\n)$ as $m_l:=n_{0,l}$. 
{}From these arguments, it is clear that the converse also holds: 
\begin{thm}[\cite{OCHA}]
For an OCHA $(\cH,\l,\n)$, let $(\cH_c',\l')$ denote the DGLA
$\Coder(T^c\cH_o)$.  The OCHA structure gives an $L_\infty$-morphism 
from $(\cH_c,\l)$ to $(\cH_c',\l')$. 
Conversely, if there exists an $L_\infty$-algebra $(\cH_c,\l)$ and 
an $L_\infty$-morphism 
from it to the DGLA  $(\cH_c',\l')$ of an $A_\infty$-algebra 
$(\cH_o,\m)$, then 
one obtains an OCHA. 
 \label{thm:deform}
\end{thm}

For an $L_\infty$-algebra $(\cH_c,\l)$, 
let us denote by $\cMC(\cH_c,\l)$ the {\em solution space} 
of the Maurer-Cartan equation 
\begin{equation*}
 \cMC(\cH_c,\l)=\big\{\cb\in\cH_c^0\ \big|\ 
 0=\sum_{k\ge 1}\ov{k!}l_k(\cb,\cdots,\cb)\big\}\ ,
\end{equation*}
where $\cH_c^0$ is the degree zero sub-vector space of $\cH_c$. 
In addition, we have an equivalence relation $\sim$ called 
{\it gauge equivalence} 
between the solutions of the Maurer-Cartan equation. 
The moduli space of the solution space of the Maurer-Cartan equation 
for $(\cH_c,\l)$ is defined by 
\begin{equation}
 \cM(\cH_c,\l)= \cMC(\cH_c,\l) / \sim\ .
\end{equation}
Suppose that we have an $L_\infty$-morphism $\f:(\cH_c,\l)\raw (\cH_c',\l')$ 
between two $L_\infty$-algebras. 
Then it is known that the $L_\infty$-morphism $\f$ induces a map 
$\f: \cM(\cH_c,\l)\raw \cM(\cH_c',\l')$, and furthermore it is an 
isomorphism if $\f$ is an $L_\infty$-quasi-isomorphism (cf. \cite{Ko1}). 
Similar facts hold also for $A_\infty$-algebras 
and also for OCHAs \cite{OCHA}, 
but what is relevant here is just the $L_\infty$ case.

Note that, for an $L_\infty$-algebra (DGLA) 
$(\cH_c',\l')=\Coder(T^c\cH_o)$ as above, 
its moduli space 
$\cM(\cH_c',\l')$ is the moduli space of deformations 
of the $A_\infty$-algebra 
in the space of weak $A_\infty$-algebras. 
Thus, we have the following: 
\begin{cor}[$A_\infty$-structure parameterized by the moduli space 
of $L_\infty$-structures \cite{OCHA}]
For an $L_\infty$-algebra $(\cH_c,\l)$ and an $A_\infty$-algebra 
$(\cH_o,\m)$, 
suppose there exists 
an OCHA $(\cH=\cH_c\oplus\cH_o,\l,\n)$ such 
that $(\cH_o,\{n_{0,k}\})=(\cH_o,\m)$. 
Then, for each element $c\in\cM(\cH_c,\l)$, 
we have a weak $A_\infty$-algebra which is 
a deformation of the original $A_\infty$-algebra $(\cH_o,\m)$. 
If $\{n_{1,k}\}_{k\ge 0}:\cH_c\raw \cH_c'$ gives a quasi-isomorphism 
of complexes $(\cH_c,l_1)\raw(\cH_c',l'_1)$, 
then all the equivalence classes of deformations of $(\cH,\m)$ as
weak $A_\infty$-algebras, described by $\cM(\cH_c',l'_1)$, 
are in one-to-one correspondence with the space $\cM(\cH_c,\l)$. 
 \label{cor:Ainfty}
\end{cor}

\subsection{The construction of deformation quantization by Kontsevich}
\label{ssec:K}

The deformation quantization problem is 
to construct a star-product corresponding to the Poisson algebra 
on a manifold $M$. 
Namely, 
for a formal (deformation) parameter $\nu$ and a given Poisson algebra 
$(\cA=C^\infty(M), \cdot, \{\ ,\ \})$, 
a bilinear, bidifferential $\nu$-linear map 
$ *:\cA[[\nu]]\otimes\cA[[\nu]]\raw\cA[[\nu]]$, 
\begin{equation}
 f*g = \sum_{r=0}^\infty m_r (f,g) \nu^r\ ,\qquad f,g\in\cA[[\nu]]
 \label{star}
\end{equation}
is called a {\em deformation quantization} of $M$ if 
$m_0(f,g)=f\cdot g$, the usual commutative product on 
$C^\infty(M)$, 
$m_1(f,g)=\half\{f,g\}$ and 
the {\em star product} $*$ is associative \cite{BFFLS}. 
(Notice here $m_r$ is still a function
of two variables and should not be confused with the $m_l$
of an $A_\infty$-algebra.)
In \cite{Ko1}, Kontsevich reformulated this problem in a 
homotopy algebraic set-up. For any associative algebra $\cal A$,
deformations as associative multiplications are controlled by
the Hochschild complex, which is essentially 
$\Coder(T^c {\cal A})$ and hence a DGLA \cite{gerst}. 
In fact, control is equally well exercised by any quasi-isomorphic
DGLA or even $L_\infty$-algebra \cite{SS}. The obstructions to existence
and to equivalence are identified by the quasi-isomorphism. For
the special case of ${\cal A}=C^\infty(M)$, the deformations relevant
to deformation quantization are controlled by the sub-complex 
of multi-differential Hochschild cochains, 
which we denote $D_{poly}(M)$ (Definition \ref{defn:Dpoly}) 
and which is quasi-isomorphic to the full Hochschild complex.
%
%AAA
%For $M={\mathbb R}^n$, t
The smooth analog 
of the Hochschild-Kostant-Rosenberg theorem \cite{HKR} 
(an explicit proof can be found in \cite{GR}) equates the 
Hochschild cohomology with $T_{poly}(M)$, 
the space of polyvector fields, which possesses a DGLA structures 
with $d=0$ and the Schouten-Nijenhuis bracket.  

Kontsevich treated these DGLAs as $L_\infty$-algebras and 
obtained 
the existence and the classification of deformation quantizations 
by constructing an $L_\infty$-morphism 
between $T_{poly}(M)$ and $D_{poly}(M)$, 
which is in a sense a non-linear generalization of a DGLA map. Moreover,
his specific $L_\infty$-morphism  provides a specific star product. 
In this setting, the space of Poisson structures 
and the space of star-products 
given by bi-differential operators are then described 
by the Maurer-Cartan equations of the corresponding DGLAs. 

In this subsection, we shall first present these tools 
and then relate them to an OCHA. 
\begin{defn}[DGLA of multi-differential operators for 
$C^\infty(M)$]
For $\cA = C^\infty (M)$, 
denote by $D_{poly}^k(M)$ the space of multi-linear maps from 
$\cA^{\otimes (k+1)}$ to $\cA$ of multi-differential operators. 
Then define $D_{poly}=\bigoplus_{k \in \Z}D_{poly}^k$ by 
\begin{equation*}
 D_{poly}^k=D_{poly}^k(M)\ (k\ge -1)\ ,\qquad D_{poly}^k=0\ (k<-1)\ .
\end{equation*}
For $d$, we take the Hochschild coboundary operator. Namely, 
for any $C\in D_{poly}^k$, $d$ is given by 
\begin{align*}
(dC) ( g_0 , \cdots , g_{k+1} )& = g_0 C( g_1 , \cdots , g_{k+1}
) - \sum_{r=0}^k (-1)^r C( g_0 , \cdots , g_r g_{r+1} , \cdots ,
g_{k+1} )  \\
&\qquad \qquad + (-1)^k C( g_0 , \cdots , g_k ) g_{k+1}\ .
\end{align*}
We take for $[\ ,\ ]$ the Gerstenhaber bracket \cite{gerst}. 
Namely, for $C\in D_{poly}^k$ , $C'\in D_{poly}^{k'}$, it is defined by 
\begin{equation*}
 \begin{split}
 &[C,C']=C\circ C'-(-1)^{k k'}C'\circ C \\
 &C\circ C'( g_0 ,\cdots ,g_{k + k'})
 \ \ =\sum_{r=0}^{k}(-1)^{r k'}C( g_0,\cdots,g_{r-1}, C'
 (g_r,\cdots,g_{r+k'}),g_{r+k'+1},\cdots,g_{k + k'}) .
 \end{split}
\end{equation*}
Then $(D_{poly},[\  ,\  ],d )$ forms a DGLA. 
 \label{defn:Dpoly}
\end{defn}
One can see that $d$ can be written as 
\begin{equation*}
 (-1)^k(dC) ( g_0 , \cdots , g_{k+1} )= [\mb, C]\ ,\qquad C\in D_{poly}^k\ ,
\end{equation*}
where $\mb$ is the usual commutative product of functions $\mb(f,g)=f\cdot g$. 
Though the operation $[\mb,\ ]$ is different in sign from the original $d$, 
$[\mb,\ ]$ also forms a DGLA on $D_{poly}$ with $[\ ,\ ]$. 
So, we take this as $d$. 
This DGLA is described as a sub-DGLA of $(\Hom,d, [\ \ ,\ \ ])$. 
First, set $\cH_o^{-1}=\cA$ and $\cH_o^k=0$ otherwise. 
Then $C:\cA^{\otimes (k+1)}\raw\cA$ has degree $k$ as defined above. 
Namely, $D_{poly}(M)$ is included in the restricted sub-vector space  
$\Hom_{sub}=\oplus_k \Hom_k^{k-1}$ of $\Hom$.

Let us state the necessary condition for the existence of 
a deformation quantization in the language of DGLAs here. 
If we write the star product (\ref{star}) as $f*g=m(f,g)$, $m\in \Hom_2^1$, 
the associativity condition $(f*g)*h=f*(g*h)$ is expressed 
algebraically as $[m, m]=0$. 
This just indicates that $m$ defines a codifferential on $T^c(\cA)$ 
as previously or, equivalently, that $m$ is associative. 
Since $m$ should be obtained as a deformation of $\mb$, writing
$m=\mb+\theta$, $\theta\in\nu D_{poly}[[\nu]]$, 
one gets a Maurer-Cartan equation in the 
DGLA $(\nu D_{poly}[[\nu]],d,[\ ,\ ])$, 
\begin{equation}
 d\theta + \half[\theta,\theta]=0\ . \label{DGLAmceq}
\end{equation}
\begin{defn}[DGLA of Polyvector fields]
For $k \geq -1$, set 
$T_{poly}^k(M):=\Gamma(M ,\wedge^{k+1} TM )$, and define 
$T_{poly}=\bigoplus_k T_{poly}^k$ by 
\begin{equation*}
 T_{poly}^k=T_{poly}^k(M)\ (k\ge -1)\ ,\qquad T_{poly}^k=0\ (k<-1)\ .
\end{equation*}
Here, when $k=-1$ we set $T_{poly}^{-1}=T_{poly}^{-1}(M)=C^\infty(M)$. 
The differential $d$ is defined by $d=0$. Therefore, 
the cohomology of the complex of $T_{poly}^n$ with respect to $d$ 
coincides with $T_{poly}^n$ itself. 
$[\ ,\ ]$ is taken to be the {\em Schouten-Nijenhuis bracket} \cite{Sc,Ni}. 
For $\xi_s,\eta_t\in T_{poly}^0=\Gamma(M, TM)$, 
the bracket of $\xi_0\wedge\cdots\wedge\xi_k,\in T_{poly}^k$ with 
$\eta_0\wedge\cdots\wedge\eta_l\in T_{poly}^l$, $k,l \geq 0$, 
is defined by 
\begin{align*}
&[\xi_0\wedge\cdots\wedge\xi_k,\eta_0\wedge\cdots\wedge\eta_l]=\\
&\sum_{i=0}^k \sum_{j=0}^l (-1)^{s+t} [ \xi_s , \eta_t ] \wedge
\xi_0 \wedge \cdots \wedge \xi_{s-1} \wedge \xi_{s+1} \wedge \cdots
\wedge \xi_k \wedge \eta_0 \wedge \cdots \wedge \eta_{t-1} \wedge
\eta_{t+1} \wedge \cdots \wedge \eta_l\ ,
\end{align*}
and for $k\geq 0,\ h\in T_{poly}^{-1}=C^\infty(M)$, the bracket is 
\[
[\xi_0 \wedge \cdots \wedge \xi_k , h] = \sum_{r=0}^k
(-1)^r \xi_r(h) ( \xi_0 \wedge \cdots \wedge \xi_{r-1} \wedge \xi_{r+1}
\wedge \cdots \wedge \xi_k )\ .
\]
We can define $\cH_c^{k-1}:=T_{poly}^k$  
with this Schouten-Nijenhuis bracket.

Now, a bivector $\alpha=\sum_{i,j}\alpha^{ij}
(\partial/\partial {x_i})\wedge(\partial/\partial {x_j})
\in T^1_{poly}$ in the local expression 
represents the Poisson bracket by 
$\{f, g\}=\sum_{i,j}\alpha^{ij}
(\partial {f}/\partial {x_i})(\partial {g}/\partial {x_j})$, 
where $(x_1,\cdots,x_n)$ are local coordinates of $M$. 
This bracket by definition satisfies 
all the axioms of a Poisson algebra except the Jacobi identity. 
The Jacobi identity is then described by  
\begin{equation}
[ \alpha ,\alpha ]=0\ .\label{dgLajacobi}
\end{equation}
Since $d=0$, eq.(\ref{dgLajacobi}) is also the Maurer-Cartan equation. 
A bivector satisfying (\ref{dgLajacobi}) is called a 
Poisson bivector. 
Similarly, a quantum deformation of the Poisson 
bivector is defined by $\alpha_{[\nu]}\in \nu T^1_{poly}[[\nu]]$ 
satisfying 
\begin{equation}
 [\alpha_{[\nu]},\alpha_{[\nu]}]=0\ .
 \label{qpoisson}
\end{equation}
In the expansion $\alpha_{[\nu]}=\nu\alpha_1+\nu^2\alpha_2+\cdots$, 
the original classical Poisson bivector $\alpha$ is $\alpha_1$. 
In fact, expanding (\ref{qpoisson}) in terms of powers of 
$\nu$, 
one can see that the lowest identity reads $[\alpha_1, \alpha_1]=0$. 
 \label{defn:Tpoly}
\end{defn}
Thus, the conditions that an element in 
$\nu T_{poly}^1[[\nu]]$ is a Poisson bracket 
and that an associative product on $C^\infty(M)[[\nu]]$ as an element
in $\nu D_{poly}^1[[\nu]]$ 
are both described by Maurer-Cartan equations for DGLAs.
Let us set two $L_\infty$-algebras $(\cH_c,\l)$ and $(\cH_c',\l')$ 
as the suspension of DGLAs $T_{poly}$ and $D_{poly}$, 
respectively. 
Namely, we have $\cH_c^k=T_{poly}^{k+1}$ and 
${\cH_c'}^k=D_{poly}^{k+1}$. 
The Maurer-Cartan equations for DGLAs
(\ref{qpoisson}),(\ref{DGLAmceq}) are of course rewritten as 
the Maurer-Cartan equations for $L_\infty$-algebras 
through the suspension in Remark \ref{rem:s}. 

The existence and the classification of the deformation quantization 
follows from the 
formality theorem \cite{Ko1}, which claims  
the existence of 
an $L_\infty$-quasi-isomorphism $\f: (\cH_c,\l)\raw (\cH_c',\l')$. 
Note that this fact implies that 
the DGLA of Hochschild cochains is (homotopically) formal; 
in particular, the higher $L_\infty$-structures vanish on its cohomologies. 
In order that $\f$ is an $L_\infty$-quasi-isomorphism, the chain map 
$f_1:(\cH_c,d_c=0)\raw (\cH_c',d_c')$ 
must be a quasi-isomorphism, 
that is, $f_1$ must induce an isomorphism on cohomologies. 
One may set 
\begin{equation*}
\( f_1(\xi_{i_1}\wedge\cdots\wedge\xi_{i_k})\)\, (g_1,\cdots,g_k)
=\ov{k!}\sum_{\sigma\in\S}(-1)^{\epsilon(\sigma)}
(\xi_{i_{\sigma(1)}}g_1)\cdots (\xi_{i_{\sigma(k)}}g_k) \ .
\end{equation*}
Kontsevich constructs all the higher multi-linear maps $f_k$ explicitly 
as  local expressions on $M=\R^n$ in terms of Feynman graphs, 
which are just those derived from 
a certain topological open-closed string theory as 
revealed explicitly by Cattaneo-Felder \cite{CF1} 
(see for review \cite{MaKa}).

When an $L_\infty$-quasi-isomorphism $\f=\{f_1,f_2,\cdots\}$ is given, 
it preserves the Maurer-Cartan equations, 
and the deformed Poisson bivector is given by 
the following `nonlinear map': 
\begin{equation}
 \theta=\sum_{k=1}^{\infty}\ov{k!}
f_k(\alpha_{[\nu]},\cdots,\alpha_{[\nu]})\ ,
 \label{alphatdef}
\end{equation}
for $\alpha_{[\nu]}\in\cM(\cH_c,\l)$ and 
$\theta\in\cM(\cH_c',\l')$. 
Here the $L_\infty$-quasi-isomorphism $\f: (\cH_c,\l)\raw (\cH_c',\l')$
has been extended by tensoring with the formal power series
$\nu\C[[\nu]]$. 
If we expand the deformation 
$m=\mb+\theta, \theta\in \nu D_{poly}[[\nu]]$ as in eq.(\ref{star}): 
$$m=\mb+\nu m_1+\nu^2m _2+\cdots$$ with $m_1 = (1/2)\alpha_1$, 
Kontsevich's quasi-isomorphism of $L_\infty$ algebras then
provides a choice for $m_2$ and in fact for all the $m_i$.

Now, let us summarize Kontsevich's deformation quantization 
in terms of OCHAs. 
First, we set $\cH^{r-2}_c:=\Gamma(M, \wedge^r TM)$. 
It forms a formal $L_\infty$-algebra 
with $l_2$ the Schouten-Nijenhuis bracket, and $l_1=l_3=\cdots=0$. 
For $\cH_o$, we take $\cH_o^{-1}:=C^\infty(M)$ and $\cH_o^k=0$ for $k\ne -1$. 
The $A_\infty$-structure is $n_{0,2}=\mb$, 
the usual commutative product of functions
$C^\infty(M)$, and $n_{0,l}=0$ except for $l=2$. 
The multi-linear maps of the $L_\infty$-quasi-isomorphism 
are then identified as the adjoints 
$\{n_{k,l}:(\cH_c)^{\otimes k}\otimes(\cH_o)^{\otimes l}\raw\cH_o\}$ 
with $k\ge 1$. 
In particular, $n_{1,2}(\alpha_{[\nu]};f,g)=(1/2)\{f, g\}$, and 
$n_{1, k}(\xi_{i_1}\wedge\cdots\wedge\xi_{i_k}; g_1,\cdots,g_k)=(1/k!)
\sum_{\sigma\in\S}(-1)^{\epsilon(\sigma)}
(\xi_{i_{\sigma(1)}}g_1)\cdots (\xi_{i_{\sigma(k)}}g_k)$. 
These structures form a minimal OCHA on cohomology. 
Corollary \ref{cor:Ainfty} then implies that, 
for a fixed element of $\cMC(\cH_c,\l)$, an $A_\infty$-structure 
is obtained. 
However, in this situation, 
since $\Hom$ is restricted to $\Hom_{sub}=\oplus \Hom_r^{r-1}$ and 
the elements of Maurer-Cartan equations, 
especially $\cM(\cH_c',\l')$ are degree zero, 
the deformed $A_\infty$-structure also has $m_2\in (\Hom_2[1])^0$ only, 
\ie, $m_1$ and higher product $m_3,m_4,\cdots$ are absent. 
Equivalently, 
when $\cH_c$ is restricted to its degree zero part $\cH_c^0$, 
$n_{k,l}$ vanishes except for $(k,l)=(k,2)$. 
This $m_2$ is just the star-product, 
an associative but noncommutative product 
$C^\infty(M)\otimes C^\infty(M)\raw C^\infty(M)$ 
of a deformation quantization. 
The next example below is a natural extension of this situation, 
but in the case that $\Hom$ is not restricted to $\Hom_{sub}$.

 \subsection{Open-closed B-model}
\label{ssec:Bmodel}

\def \Mh{{\hat M}}

A natural extension of Kontsevich's deformation quantization set-up is to
the B-model side of homological mirror symmetry \cite{mirror}. 

The mirror symmetry, a symmetry between Calabi-Yau manifolds, 
can be interpreted as topological closed string physics. 
There are two types of topological string theories 
whose target spaces are Calabi-Yau manifolds. 
One is called the A-model, which depends only on the complexified 
symplectic structure and is independent of the complex structure 
of the Calabi-Yau manifold. 
Another one, the B-model in contrast depends only on the complex
structure. 
For a given Calabi-Yau manifold $M$, the mirror symmetry conjecture 
is the existence of a mirror Calabi-Yau manifold $\Mh$ 
such that the A-model closed string on $\Mh$ is equivalent to 
the B-model closed string on $M$ and vice versa \cite{W1}. 
Homological mirror symmetry is thought of as an open string version \cite{W2} 
of the mirror symmetry conjecture. 
Open string theory in general includes some kind of D-branes, 
which form a D-brane category (see \cite{Dbcat}); 
the D-branes and open strings are identified with objects and morphisms 
between the objects. 
For the tree open string A-model, the corresponding category is 
Fukaya's $A_\infty$-category \cite{Fukaya}, 
which depends only on the complexified symplectic structure. 
On the other hand, what is constructed on the B-model side is 
a category of holomorphic vector bundles or coherent sheaves 
more generally. 
The homological mirror symmetry conjecture \cite{mirror} 
then states that 
the Fukaya category on a Calabi-Yau manifold $M$ is in some sense 
equivalent to the category of coherent sheaves on the mirror dual 
Calabi-Yau manifold $\Mh$. 
Now, the conjecture is checked successfully 
in the case $M$ is an elliptic curve \cite{mirror,PZ,Poli}, 
an abelian variety \cite{F,KoSo}, a quartic surface \cite{Sei}, 
and so on. For noncommutative two-tori, see \cite{foliation, KimKim,fol2} 
and a related work \cite{PoSc}. 

One of the original motivations to argue for
this homological mirror symmetry conjecture \cite{mirror} is 
that it might explain the (tree closed string parts of) mirror symmetry: 
the family (deformations) of tree open string A (resp. B)-models should be 
in one-to-one correspondence with that of 
tree closed string A (resp. B)-model, and 
the tree closed string structure should follow from the corresponding 
family of the tree open string structures. 
These concepts have their background 
in the open-closed string physics in our sense. 
Since the A-side and B-side should be mirror dual, 
they should have isomorphic structures 
in some sense. However, that which is directly related to us 
is the B-model side, 
since there the classical solution of string world sheet theory is  only a 
constant map (no world sheet instanton) and 
the corresponding moduli spaces are just the usual ones 
of Riemann surfaces with boundaries and punctures \cite{W1}. 

One can see that the tree open-closed B-model is just a particular example 
of the arguments in subsection \ref{ssec:def}, 
and gives a natural extension of that in the previous subsection. 
However, the mathematical formulation of the 
tree open-closed B-model is not yet established completely, 
nor is there known the explicit formula 
as in the case of deformation quantization 
in the previous subsection. 

An interesting attempt and a partial 
result can be found in Hofman's work \cite{Hof2}. 
We can identify some set of multi-linear maps $\{l_k, n_{p,q}\}$ 
on open/closed string observables 
for this situation. 
Of course we could have an infinite number of homotopy equivalent open-closed 
homotopy structures. 
For instance, for the tree closed string part, the world sheet action of 
the B-twisted topological string theory as given in \cite{W1} has
the space of observables  which is identified with the cohomology of 
$\oplus_{p,q}\Gamma(M,\wedge^p TM\otimes\wedge^q\overline{T^*M})$ 
with respect to the Dolbeault operator $\bpart$ 
and hence, in principle, one can compute closed string $(k+1)$-point functions 
%(indirectly, see the footnote below) 
(= scattering amplitudes) related to $l_k$. 
However, in general this is complicated; it is better if one can find 
a corresponding string field theory in a simple form. Such a string
field theory is  given by \cite{BCOV}, 
where the B-twisted topological closed string field theory action 
consists of the kinetic term and a three point vertex only, 
and so a DGLA structure is associated to it. 
Note that the equations of motions are just the Maurer-Cartan equations 
defining deformations of the complex structures, with additional 
extended directions mentioned below. 
The string field theory gives, 
at least for tree level (genus zero), a simple way of calculating 
$(k+1)$-point functions in terms of Feynman rules, 
and this procedure just coincides with taking 
a minimal model of the DGLA from a homotopy algebraic point of view. 
\footnote{However, for the action given in \cite{BCOV}, 
they use an inner product 
which is not nondegenerate in our sense. 
In particular, it vanishes on the cohomology. 
Together with an additional structure called 
a differential Gerstenhaber structure, 
the pull-back of this action with respect to 
the $L_\infty$-quasi-isomorphism, 
or equivalently the superpotential or the collection of 
tree closed string amplitudes, has a Frobenius structure \cite{BCOV,BK}, 
even though the minimal $L_\infty$-structure is trivial, 
all $l_k=0$. 
We do not deal with this Frobenius structure in this paper. }
 
In a similar way, one can also consider open strings 
in the B-twisted topological string theory, and, in a similar spirit, 
one can construct a particular 
open string topological string field theory action 
\cite{W2, BCOV}, called a holomorphic Chern-Simons action, 
a holomorphic version of the usual Chern-Simons action 
or Witten's bosonic open string field theory \cite{W}. 
Thus, it has a structure of a differential graded associative algebra 
(DGA) with cyclicity, a cyclic $A_\infty$-structure. 
Again, tree open string world sheet scattering amplitudes are obtained 
by taking the minimal model of the DGA. 

Although the B-twisted topological string is controlled 
by such a simple DGLA or DGA 
for the closed or open string case, respectively, 
it is known that the same story does not hold 
for tree open-closed string in general \cite{Z2}. 
Therefore, we consider here a minimal OCHA structure 
whose purely closed string part $\{l_k\}$ 
and purely open string part $\{m_k=n_{0,k}\}$ 
are given by taking the minimal models of the DGLA and DGA, respectively. 
For the closed string side, the corresponding DGLA structure is 
$(\Gamma(M,\wedge^\star TM\otimes\wedge^\star\overline{T^*M}), 
\bpart,[\ ,\ ])$, 
where $[\ ,\ ]$ is the Schouten-Nijenhuis bracket, 
the one in Definition \ref{defn:Tpoly} extended 
to $\overline{T^*M}$ naturally.
However, for closed strings, the $\partial\bpart$-lemma 
and Tian-Todorov's lemma 
lead to the corresponding minimal $L_\infty$-structure, 
given by the procedure in section \ref{sec:MMth}, 
being trivial (see \cite{BCOV,BK}). 
Thus, we have $l_k=0$ for $k\ge 1$ 
for $\cH_c^k=\oplus_{p+q-2=k}H^q(\wedge^p TM)$, 
and the corresponding moduli space is itself : 
\begin{equation*}
 \cM(\cH_c, \l)=\cH_c^0\ .
\end{equation*} 
Here, we restrict $\cH_c$ to its degree zero part only. 
It might be reasonable to think 
that one can deform in these $\cH_c^0$ directions finitely in principle
and that the other $\cH_c^k$ directions provide fibers 
of an infinitesimal neighborhood. 
In fact, the whole deformation space $\cH_c$ is called the 
extended moduli space in Barannikov-Kontsevich \cite{BK}. 
Note that $H^1(\wedge^1 TM)\in\cH_c^0$ describes 
the complex structure deformations. 
This is the original deformation theory of complex structures, 
and the Barannikov-Kontsevich's 
set-up can be thought of as an extension of it. 

Next, for pure open string structure, 
we should stress what is taken for $\cH_o$. 
The open string theory forms a D-brane category, 
which should be treated as an $A_\infty$-category in our context. 
The objects, B-type D-branes, are the coherent sheaves on $M$. 
Thus, $\cH_o$ is identified with the space of morphisms between them. 
A general construction of a minimal model in this situation is 
found in \cite{Laz, LazR, Tom, Dbcat}, and see \cite{fol2} 
for an explicit construction in the noncommutative two-tori case. 
However, for simplicity, here we consider the case that 
the object is only the structure sheaf $\cO(M)$. 
One can see that this simplified situation is enough for 
our purpose here 
under some appropriate assumptions. 
The differential is then simply $\bpart$, and 
one obtains a minimal model of 
DGA $(\cO^{0,*}(M), \bpart, \wedge )$, 
which we denote by $(\cH_o, \mb=\{n_{0,k}\}_{k\ge 2})$, where 
$\cH_o^{k-1}=H^k(\cO^{0,*}(M))$. 

For this particular choice of $\cH_o$, 
let us consider the space of multi-linear maps
$\Hom:=\oplus \Hom_k^r$, 
\begin{equation*}
 \Hom_k^r=\Hom^r(\cH_o^{\otimes k},\cH_o)\ .
\end{equation*}
Again, the result of Gerstenhaber-Schack \cite{gerst-schack}, 
in a similar way as the Hochschild-Konstant-Rosenberg theorem 
does in the previous subsection, implies that the cohomology of 
$\Hom=\oplus_{k,r}\Hom_k^r$ coincides with $\cH_c$ itself 
(see \cite{mirror}).

Alternatively, 
the existence of an $L_\infty$-quasi-isomorphism 
from $(\cH_c,\l)$ to $(\cH_c',\l')$ 
is guaranteed at least physically, 
since $\{n_{k,l}\}$ can be constructed as 
open-closed disk amplitudes of the open-closed B-model, 
where the string world sheet action is of the same form 
as the one for pure closed strings, 
and the space of observables are just those used in separate
open/closed B-model. 
An open-closed disk amplitude is then obtained by 
the integral of a disk correlation function, calculated 
in the usual way in physics, over the moduli space of the 
corresponding disk with punctures, both in the interior and on the
boundary. 

Moreover, it was checked in \cite{Hof2} that $n_{1,q}$ given 
by this physical argument 
in fact gives the linear part of the $L_\infty$-quasi-isomorphism $f_1$
\begin{equation*}
 n_{1, k}(\xi_{i_1}\wedge\cdots\wedge\xi_{i_k}, f_1,\cdots,f_k)
 =\ov{k!}\sum_{\sigma\in\S}(-1)^{\epsilon(\sigma)}
 (\xi_{i_{\sigma(1)}}f_1)\cdots (\xi_{i_{\sigma(k)}}f_k)
\end{equation*}
where $\xi_i\in TM$ acts on $f_i\in\cH_o$ as the Lie derivative. 
Also, in the spirit of homological mirror symmetry, 
since the collection $\{n_{k,l}\}$ 
gives an $L_\infty$-quasi-isomorphism 
even with the restriction to the category of coherent sheaves 
to $\cO(M)$, it does also in case the full category 
would be treated for the open string side.

To summarize, we set $\cH_c^k=\oplus_{p+q-2=k}H^q(\wedge^p TM)$, 
$\cH_o^{k-1}=H^k(\cO^{0,*}(M))$, 
and $\Hom_k^r=\Hom^r(\cH_o^{\otimes k},\cH_o)$. 
Then we have $l_k=0$ on $\cH_c$ and $n_{0,q}$ is 
the minimal $A_\infty$-structure of DGA $(\cO^{0,*}(M), \bpart, \wedge)$. 
The operation $l'$ is defined 
as in subsection \ref{ssec:def}, 
that is, $l_1'=[\mb,\ ]$ with $\mb=\sum_{0,q}n_{0,q}$ and $\l'_2$ 
is related to the commutator 
$[\ \, ,\ \, ]$ through the suspension. 

The corresponding OCHA structure in this case reduces to 
the generalized WDVV relation discussed in \cite{HLL}. 
Namely, it corresponds to a minimal OCHA 
with trivial $L_\infty$-structure. 
At present, the explicit form of multi-linear maps 
$n_{p,q}$ of $p\ge 2$ and $q\ge 0$ are not known in general, 
though physically existence is guaranteed by the open-closed 
scattering amplitudes of disks in B-twisted topological string theory. 
There is an interesting restriction where we can understand 
the form of $n_{k,l}$. 
Hofman discussed in \cite{Hof2} that if we restrict $\cH_c$ to $\wedge^p TM$, 
the situation reduces to a complex version 
of  Kontsevich's deformation quantization 
and hence $n_{p,q}$ are obtained in a similar way. 

An interesting difference from the deformation quantization set-up 
in the previous subsection is that, 
since $\Hom=\oplus \Hom_k^r$ is not restricted to $\Hom_{sub}$, even if $r=1$ 
we can have nontrivial deformed $m_k$. 
In particular, we could in general obtain a deformation 
of the $A_\infty$-structure $\mb$ to a weak $A_\infty$-structure, 
which should be one of the interesting future directions to 
be investigated. 

In this case, for the closed string part, 
the $L_\infty$-structure is trivial, including the bracket, and 
hence the obstructions vanish.  
In a more general model, however, 
it should not be trivial, as discussed by 
Huebschmann-Stasheff \cite{jh-jds}. 
It should be interesting to find such models which can be 
calculated explicitly.

\begin{center}
\noindent{\large \textbf{Acknowledgments}}
\end{center}

We would like to thank T.~Kimura, E.~Harrelson, M.~Markl and A.~Voronov 
for valuable discussions.


\begin{thebibliography}{99}


%\cite{Alexandrov:1997kv}
\bibitem{AKSZ}
M.~Alexandrov, M.~Kontsevich, A.~Schwartz and O.~Zaboronsky,
``The Geometry of the master equation and topological quantum field theory,''
Int.\ J.\ Mod.\ Phys.\ A {\bf 12} (1997) 1405, 
hep-th/9502010.
%%CITATION = HEP-TH 9502010;%%



\bibitem{BK}
S.~Barannikov and M.~Kontsevich, 
``Frobenius manifolds and formality of Lie algebras of polyvector fields,'' 
Int. Math. Res. Notices 1998, %no. 4, 
201--215. 



%\cite{Batalin:1981jr}
\bibitem{BV1}
I.~A.~Batalin and G.~A.~Vilkovisky,
``Gauge algebra and quantization,''
Phys.\ Lett.\ B {\bf 102} (1981) 27.
%%CITATION = PHLTA,B102,27;%%



%\cite{Batalin:1983jr}
\bibitem{BV2}
I.~A.~Batalin and G.~A.~Vilkovisky,
``Quantization of gauge theories with linearly dependent generators,''
Phys.\ Rev.\ D {\bf 28} (1983) 2567, 
[Erratum-ibid.\ D {\bf 30} (1983) 508].
%%CITATION = PHRVA,D28,2567;%%



\bibitem{BFFLS}
F.~Bayen, M.~Flato, C.~Fr{\o}nsdal, A.~Lichnerowicz and D.~Sternheimer,
``Deformation theory and quantization I, II,'' 
Ann. Phys. {\bf 111} (1978), 61-110, 111-151. 



\bibitem{BCOV}
M.~Bershadsky, S.~Cecotti, H.~Ooguri and C.~Vafa, 
``Kodaira-Spencer theory of gravity and exact results for quantum string amplitudes,''
Comm. Math. Phys. {\bf 165} (1994), %no. 2, 
311--427.



\bibitem{BoVo} 
J.~M.~Boardman and R.~M.~Vogt,
{\em Homotopy Invariant Algebraic Structures on Topological
Spaces}, Lecture Notes in Mathematics {\bf 347}, Springer, 1973.



\bibitem{CF1} 
A.~S.~Cattaneo and G.~Felder, 
``A path integral approach to the Kontsevich quantization formula,''
Commun. Math. Phys. {\bf 212} (2000) 591--611, 
math.QA/9902090. 



\bibitem{Fukaya}
K.~Fukaya, 
``Morse homotopy, $A\sp \infty$-category, and Floer homologies,'' 
Proceedings of GARC Workshop on Geometry and Topology
'93 (Seoul, 1993), pp.1--102, Lecture Notes Ser., 18, 
Seoul Nat. Univ., Seoul, 1993. 



\bibitem{F}
K.~Fukaya, 
``Mirror symmetry of abelian varieties and multi theta functions,''
J. Algebraic Geom. {\bf 11} (2002), %no. 3, 
393--512, 
preprint, Kyoto University, 1998.



%\cite{Gaberdiel:1997ia}
\bibitem{GZ}
M.~R.~Gaberdiel and B.~Zwiebach,
``Tensor constructions of open string theories I: Foundations,''
Nucl.\ Phys.\ B {\bf 505} (1997) 569, 
hep-th/9705038.
%%CITATION = HEP-TH 9705038;%%



\bibitem{gerst}
M.~Gerstenhaber, 
%\bibitem{gerst:coh}
``The cohomology structure of an associative ring,''
Ann. Math. %(2) 
{\bf 78} (1963) 267--288 ; 
``On the deformation of rings and algebras,''
Ann. Math. {\bf 79} (1964) 59--103.



\bibitem{gerst-schack}
M.~Gerstenhaber and S.~D.~Schack, 
``Algebraic cohomology and deformation theory,'' 
Deformation theory of algebras and structures 
and applications (Il Ciocco, 1986), pp.11--264, 
NATO Adv. Sci. Inst. Ser. C Math. Phys. Sci., {\bf 247}, 
Kluwer Acad. Publ., Dordrecht, 1988. 



%\cite{Getzler:1994yd}
\bibitem{get1}
E.~Getzler,
``Batalin-Vilkovisky algebras and 
two-dimensional topological field theories,''
Commun.\ Math.\ Phys.\  {\bf 159} (1994) 265, 
hep-th/9212043.
%%CITATION = HEP-TH 9212043;%%



\bibitem{GJ}
E.~Getzler and J.D.S.~Jones, 
``$A_{\infty}$-algebras and the cyclic bar complex,'' 
Ill. J. Math. {\bf 34}, 256 (1990).



\bibitem{GeK1}
E.~Getzler and M.~M.~Kapranov, 
``Cyclic operads and cyclic homology,''
Geometry, topology, $\&$ physics, pp.167--201, Conf. Proc. Lecture 
Notes Geom. Topology, IV, Internat. Press, Cambridge, MA, 1995. 



\bibitem{gugen}
V.~K.~A.~M.~Gugenheim, 
``On a perturbation theory for the homology of the loop-space,''
J. Pure Appl. Algebra {\bf 25} (1982), %no. 2, 
197--205. 



\bibitem{gugen-lambe}
V.~K.~A.~M.~Gugenheim and L.~A.~Lambe, 
``Perturbation theory in differential homological algebra. I,''
Ill. J. Math. {\bf 33} (1989), %no. 4, 
566--582. 



\bibitem{GLS:chen}
V.~K.~A.~M.~Gugenheim, L.~A.~Lambe and J.~D.~Stasheff, 
``Algebraic aspects of Chen's twisting cochain,''
Ill. J. Math. {\bf 34} (1990), %no. 2, 
485--502. 
                                       


\bibitem{GLS}
V.~K.~A.~M.~Gugenheim, L.~A.~Lambe and J.~D.~Stasheff, 
``Perturbation theory in differential homological algebra II,''
Ill. J. Math. {\bf 35} (1991), %no. 3, 
357--373. 



\bibitem{GS}
V.~K.~A.~M.~Gugenheim and J.~D.~Stasheff, 
``On perturbations and $A\sb \infty$-structures,''
 Bull. Soc. Math. Belg. S\'er. A {\bf 38} (1986), 237--246
(1987). 



\bibitem{GR}
 S.~Gutt and J.~Rawnsley,
``Equivalence of star products on a symplectic manifold; 
 an introduction to Deligne's Cech cohomology classes,''
J. Geom. Phys.{\bf 29} (1999) 347--392.
  


\bibitem{erich}
E.~Harrelson, 
``On the homology of open/closed string theory,''
math.AT/0412249. 



%\cite{Herbst:2004jp}
\bibitem{HLL}
M.~Herbst, C.~I.~Lazaroiu and W.~Lerche,
``Superpotentials, A(infinity) relations and WDVV equations for open 
topological strings,''
JHEP {\bf 0502} (2005) 071, 
hep-th/0402110.
%%CITATION = HEP-TH 0402110;%%



\bibitem{HKR}
G.~Hochschild, B.~Kostant and A.~Rosenberg, 
``Differential forms on regular affine algebras,''
Trans. Amer. Math. Soc. {\bf 102} (1962) 383--408. 



%\cite{Hofman:2002cw}
\bibitem{Hof2}
C.~Hofman,
``On the open-closed B-model,''
JHEP {\bf 0311} (2003) 069, 
hep-th/0204157. 
%%CITATION = HEP-TH 0204157;%%



%\cite{Hofman:2000ce}
\bibitem{Hof1}
C.~Hofman and W.~K.~Ma,
``Deformations of topological open strings,''
JHEP {\bf 0101} (2001) 035, 
hep-th/0006120. 
%%CITATION = HEP-TH 0006120;%%



\bibitem{hueb-kadei}
J.~Huebschmann and T.~Kadeishvili, 
``Small models for chain algebras,'' 
Math. Z. {\bf 207} (1991), %no. 2, 
245--280. 



\bibitem{jh-jds}
J.~Huebschmann and J.~Stasheff, 
``Formal solution of the master equation via HPT and deformation theory,''
Forum Math. {\bf 14} (2002), %no. 6, 
847--868.



\bibitem{kadei1}
T.~V.~Kadeishvili, 
``The algebraic structure in the homology of an $A(\infty )$-algebra,''
 (Russian) Soobshch. Akad. Nauk Gruzin. SSR {\bf 108} (1982), %no. 2, 
249--252 (1983). 



%\cite{Kajiura:2001ng}
\bibitem{Ka}
H.~Kajiura,
``Homotopy algebra morphism and geometry of classical string field theories,''
Nucl.\ Phys.\ B {\bf 630} (2002) 361, 
hep-th/0112228.
%%CITATION = HEP-TH 0112228;%%



%\cite{Kajiura:2002vm}
\bibitem{foliation}
H.~Kajiura,
``Kronecker foliation, D1-branes and Morita equivalence of 
noncommutative two-tori,''
JHEP {\bf 0208} (2002) 050, 
hep-th/0207097. 
%%CITATION = HEP-TH 0207097;%%



\bibitem{thesis}
H.~Kajiura,
``Noncommutative homotopy algebras associated with open strings,''
thesis submitted to Graduate School of Mathematical Sciences, 
Univ. of Tokyo, 
math.QA/0306332.



%%\cite{Kajiura:2004rn}
\bibitem{fol2}
H.~Kajiura,
``Homological mirror symmetry on noncommutative two-tori,''
hep-th/0406233.
%%CITATION = HEP-TH 0406233;%%



\bibitem{OCHA}
H.~Kajiura and J.~Stasheff, 
``Homotopy algebras inspired by classical open-closed string field theory,''
Comm. Math. Phys. {\bf 263} (2006) 553--581, 
math.QA/0410291. 
%%CITATION = MATH-QA 0410291;%%



\bibitem{KaTe}
H.~Kajiura and Y.~Terashima, 
``Homotopy equivalence of $A_\infty$-morphisms and 
gauge transformations,''
preprint, 2003. 



%\cite{Kim:2003nq}
\bibitem{KimKim}
E.~Kim and H.~Kim,
 ``Moduli spaces of standard holomorphic bundles on a noncommutative 
complex torus,''
math.QA/0312228.
%%CITATION = MATH-QA 0312228;%%



%\cite{Kimura:1993ea}
\bibitem{KSV}
T.~Kimura, J.~Stasheff and A.~A.~Voronov,
``On operad structures of moduli spaces and string theory,''
Commun.\ Math.\ Phys.\  {\bf 171} (1995) 1, 
hep-th/9307114.



\bibitem{mirror}
M.~Kontsevich, 
``Homological algebra of mirror symmetry,'' 
Proceedings of the International Congress of Mathematicians, Vol. 1, 2 
(Z\"urich, 1994), pp.120--139, Birkh\"auser, Basel, 1995. 



%\cite{Kontsevich:1997vb}
\bibitem{Ko1}
M.~Kontsevich,
``Deformation quantization of Poisson manifolds,''
Lett. Math. Phys. {\bf 66} (2003), %no. 3, 
157--216, 
math.QA/9709040. 
%%CITATION = Q-ALG 9709040;%%



%\cite{Kontsevich:2000yf}
\bibitem{KoSo}
M.~Kontsevich and Y.~Soibelman,
``Homological mirror symmetry and torus fibrations,''
Symplectic geometry and mirror symmetry 
(Seoul, 2000), pp.203--263, 
World Sci. Publishing, River Edge, NJ, 2001, 
math.SG/0011041.
%%CITATION = MATH.SG 0011041;%%



\bibitem{lada-markl}                                                           
T.~Lada and M.~Markl, ``Strongly homotopy {L}ie algebras,''
Comm.~in Algebra (1995), 2147--2161, 
hep-th/9406095.



\bibitem{LS}
T.~Lada and J.~Stasheff,
``Introduction to sh Lie algebras for physicists,''
Int. J. Theor. Phys. {\bf 32} (1993) %no. 7, 
1087-1103.



%\cite{Lazaroiu:2001nm}
\bibitem{Laz}
C.~I.~Lazaroiu,
``String field theory and brane superpotentials,''
JHEP {\bf 0110} (2001) 018, 
hep-th/0107162.
%%CITATION = HEP-TH 0107162;%%



%\cite{Lazaroiu:2003md}
\bibitem{Dbcat}
C.~I.~Lazaroiu,
``D-brane categories,''
Int.\ J.\ Mod.\ Phys.\ A {\bf 18} (2003) 5299, 
hep-th/0305095.
%%CITATION = HEP-TH 0305095;%%



%\cite{Lazaroiu:2001bz}
\bibitem{LazR}
C.~I.~Lazaroiu and R.~Roiban,
``Holomorphic potentials for graded D-branes,''
JHEP {\bf 0202} (2002) 038, 
hep-th/0110288; 
%%CITATION = HEP-TH 0110288;%%
%\cite{Lazaroiu:2001qp}
%\bibitem{Lazaroiu:2001qp}
%C.~I.~Lazaroiu and R.~Roiban,
``Gauge-fixing, semiclassical approximation and potentials for graded
Chern-Simons theories,'' 
JHEP {\bf 0203} (2002) 022, 
hep-th/0112029.
%%CITATION = HEP-TH 0112029;%%



\bibitem{Le-Ha}
K.~Lef\`evre-Hasegawa, 
``Sur les $A_\infty$-cat\'egories,'' 
math.CT/0310337. 



\bibitem{MaKa}
Y.~Maeda and H.~Kajiura, 
``String theory and deformation quantization,'' 
Sugaku exposition {\bf 55}  (2003), %no. 3, 
245--265. 



\bibitem{markl:module} 
M.~Markl, 
``Models for operads,''
Commun. Algebra {\bf 24} (1996), %no. 4, 
1471--1500.



%%\cite{Markl:2001bj}
\bibitem{markl:loop}
M.~Markl,
``Loop homotopy algebras in closed string field theory,''
Commun.\ Math.\ Phys.\  {\bf 221} (2001) 367, 
hep-th/9711045.
%%%CITATION = HEP-TH 9711045;%%



\bibitem{markl:haha}
M.~Markl, 
``Homotopy algebras are homotopy algebras,''
Forum Math. {\bf 16} (2004) 129, 
math.AT/9907138.


                                                           
\bibitem{markl:transfer}
M.~Markl, 
``Transferring $A_\infty$ (strongly homotopy associative) structures,''
math.AT/0401007. 



\bibitem{MSS}
M.~Markl, S.~Shnider and J.~Stasheff, 
{\it Operads in algebra, topology and physics}, 
Mathematical Surveys and Monographs, {\bf 96}, 
American Mathematical Society, Providence, RI, 2002. x+349 pp. 



\bibitem{mer}
S.~A.~Merkulov, 
``Strong homotopy algebras of a K\"ahler manifold,'' 
Int. Math. Res. Notices 1999, %no. 3, 
153--164, 
math.AG/9809172. 



%\cite{Nakatsu:2001da}
\bibitem{N}
T.~Nakatsu,
``Classical open-string field theory: A(infinity)-algebra, 
renormalization group and boundary states,''
Nucl.\ Phys.\ B {\bf 642} (2002) 13, 
hep-th/0105272.
%%CITATION = HEP-TH 0105272;%%



\bibitem{Ni}
A.~Nijenhuis,
``Jacobi-type identities for bilinear differential 
concomitants of certain tensor fields\ I, II,''
Indag. Math. {\bf 17} (1955), 
390--397, 398--403.



\bibitem{Poli}
A.~Polishchuk, 
``Homological mirror symmetry with higher products,''
Winter School on Mirror Symmetry, 
Vector Bundles and Lagrangian Submanifolds (Cambridge, MA, 1999), 
pp.247--259, AMS/IP Stud. Adv. Math., 23, Amer. Math. Soc., 
Providence, RI, 2001, 
math.AG/9901025 ; 
%\bibitem{Poli2}
%A.~Polishchuk, 
``$A_{\infty}$-structures on an elliptic curve,''
Commun.\ Math.\ Phys.\  {\bf 247} (2004) 527, 
math.AG/0001048. 



%\cite{Polishchuk:2002sj}
\bibitem{PoSc}
A.~Polishchuk and A.~Schwarz,
``Categories of holomorphic vector bundles on noncommutative two-tori,''
Commun.\ Math.\ Phys.\  {\bf 236} (2003) 135, 
math.QA/0211262.
%%CITATION = MATH-QA 0211262;%%



\bibitem{PZ}
A.~Polishchuk and E.~Zaslow, 
``Categorical mirror symmetry: the elliptic curve,''
Adv. Theor. Math. Phys. {\bf 2} (1998), no. 2, 443--470, 
math.AG/9801119.



\bibitem{SS}
M.~Schlessinger and J.~Stasheff, 
``Deformaion theory and rational homotopy type,''
U. of North Carolina preprint, 1979; short version: 
``The Lie algebra structure of tangent cohomology and deformation theory,''
J. Pure Appl. Algebra {\bf 38} (1985), 313--322. 



 \bibitem{Sc}
 J.~A.~Schouten,
 ``On the differential operators of first order in tensor calculus,''
  Convegno Internazionale di Geometria Differenziale, Italia, 1953,
pp.1-7, Edizioni Cremonese, Roma (1954).



%\cite{Schwarz:1993nx}
\bibitem{Sch}
A.~Schwarz,
``Geometry of Batalin-Vilkovisky quantization,''
Commun.\ Math.\ Phys.\  {\bf 155} (1993) 249, 
hep-th/9205088.
%%%CITATION = HEP-TH 9205088;%%



\bibitem{Sei}
P.~Seidel, 
``Homological mirror symmetry for the quartic surface,''
math.SG/0310414. 

     

\bibitem{Sta} 
J.~D.~Stasheff, 
``On the homotopy associativity of $H$-spaces, I, II,''
Trans. Amer. Math. Soc. {\bf 108} (1963) 275, 293. 



\bibitem{jds:intrinsic}                                                        
J.~D. Stasheff, 
``The intrinsic bracket on the deformation complex 
of an associative algebra", 
J. Pure Appl. Algebra {\bf 89} (1993), 231--235, 
Festschrift in Honor of Alex Heller.
 


%\cite{Stasheff:1993ny}
\bibitem{Sta1993}
J.~Stasheff,
``Closed string field theory, 
strong homotopy Lie algebras and the operad actions 
of moduli spaces,''
hep-th/9304061. 
%%CITATION = HEP-TH 9304061;%%



\bibitem{Staconf}
J.~Stasheff, 
``From operads to "physically" inspired theories,''
Operads: Proceedings of Renaissance Conferences 
(Hartford, CT/Luminy, 1995), pp.53--81, 
Contemp. Math., {\bf 202}, Amer. Math. Soc., Providence, RI, 1997. 



%\cite{Tomasiello:2001yq}
\bibitem{Tom}
A.~Tomasiello,
``A-infinity structure and superpotentials,''
JHEP {\bf 0109} (2001) 030, 
hep-th/0107195.
%%CITATION = HEP-TH 0107195;%%



\bibitem{Vo}
A.~Voronov, 
``The Swiss-cheese operad,''
Homotopy invariant algebraic structures (Baltimore, MD, 1998), pp.365--373, 
Contemp. Math., {\bf 239}, 
Amer. Math. Soc., Providence, RI, 1999. 



%\cite{Witten:1986cc}
\bibitem{W}
E.~Witten,
``Noncommutative geometry and string field theory,''
Nucl.\ Phys.\ B {\bf 268} (1986) 253.
%CITATION = NUPHA,B268,253;%%



%\cite{Witten:1991zz}
\bibitem{W1}
E.~Witten,
``Mirror manifolds and topological field theory,''
Essays on mirror manifolds, pp.120--158, Internat. Press, Hong Kong,
1992, 
hep-th/9112056.
%%CITATION = HEP-TH 9112056;%%



%\cite{Witten:1992fb}
\bibitem{W2}
E.~Witten,
``Chern-Simons gauge theory as a string theory,''
Prog.\ Math.\  {\bf 133} (1995) 637, 
hep-th/9207094.
%CITATION = HEP-TH 9207094;%%



%\cite{Zwiebach:1993ie}
\bibitem{Z1}
B.~Zwiebach,
``Closed string field theory: Quantum action and the B-V master equation,''
Nucl.\ Phys.\ B {\bf 390} (1993) 33, 
hep-th/9206084.
%%CITATION = HEP-TH 9206084;%%



%\cite{Zwiebach:1998fe}
\bibitem{Z2}
B.~Zwiebach,
``Oriented open-closed string theory revisited,''
Annals Phys.\  {\bf 267} (1998) 193, 
hep-th/9705241.
%%CITATION = HEP-TH 9705241;%%



\end{thebibliography}
\end{document}